\documentclass[review]{elsarticle}

\usepackage{lineno,hyperref}
\usepackage{longtable}
\usepackage{acronym}
\usepackage{color}
\usepackage{rotating}
\usepackage{aas_macros}
\usepackage{xcolor}
\usepackage{multirow}
\usepackage{rotating}
\usepackage{booktabs}
\usepackage{float}
\usepackage{wasysym}
\usepackage{tabularx}

\journal{Icarus}

\biboptions{authoryear}

\hypersetup{draft}
\begin{document}

\begin{frontmatter}

\title{Non-Thermal Escape of the Martian CO$_2$ atmosphere over time: Constrained by Ar isotopes}
\author[address1]{H.I.M. Lichtenegger}
\author[address1]{S. Dyadechkin}
\author[address1]{M. Scherf\corref{mycorrespondingauthor}}
\cortext[mycorrespondingauthor]{Corresponding author}
\author[address1]{H. Lammer}
\author[address3]{R. Adam}
\author[address2]{E. Kallio}
\author[address1]{U.V. Amerstorfer}
\author[address2,address5]{R. Jarvinen}

\address[address1]{Space Research Institute, Austrian Academy of Sciences, Graz, Austria}
\address[address3]{Institute of Physics, IGAM, Karl Franzens University, Graz, Austria}
\address[address2]{Department of Electronics and Nanoengineering, School of Electrical Engineering, Aalto University, Espoo, Finland}
\address[address5]{Finnish Meteorological Institute, Helsinki, Finland}
\begin{abstract}
The ion escape of Mars' CO$_2$ atmosphere caused by its dissociation products C and O atoms is {simulated} from present time to $\sim 4.1$ billion years ago (Ga) by {numerical models of the upper atmosphere and its interaction with the solar wind}. The planetward-scattered pick-up ions are used for sputtering estimates of exospheric particles including $^{36}$Ar and $^{38}$Ar isotopes. Total ion escape, sputtering and photochemical escape rates are compared. For solar EUV fluxes $\geq$\,3 times  that of today's Sun (earlier than $\sim 2.6$ Ga) ion escape becomes the dominant atmospheric non-thermal loss process until thermal escape takes over during the pre-Noachian eon (earlier than $\sim 4.0\,-\,4.1$ Ga). If we extrapolate the total escape of CO$_2$-related dissociation products back in time until $\sim$ 4.1 Ga we obtain a {maximum} theoretical equivalent to CO$_2$ partial pressure of more than {$\sim 0.4$ bar through non-thermal escape during quiet solar wind conditions}. {However, surface-atmosphere interaction and/or extreme solar events such as frequent CMEs could have increased this value even further. By including the surface as a sink, up to 0.9\,bar, or even up to 1.8\,bar in case of hidden carbonate reservoirs, could have been present at 4.1\,Ga} The fractionation of $^{36}$Ar/$^{38}$Ar isotopes through sputtering and volcanic outgassing from its initial chondritic value of 5.3, as measured in the 4.1 billion years old Mars meteorite ALH 84001, until the present day, however, can be reproduced for assumed CO$_2$ partial pressures of {$\sim0.01 -- 0.4$\,bar without, and $\sim0.4 -- 1.8$\,bar including surface sinks, and} depending on the cessation time of the Martian dynamo (assumed between 3.6\,--\,4.0 Ga) - if atmospheric sputtering of Ar started afterwards.
\end{abstract}

\begin{keyword}
atmospheric loss \sep hybrid simulation \sep atmospheric sputtering \sep hot particles
\end{keyword}

\end{frontmatter}

\section{Introduction}
The evolution and escape of Mars' CO$_2$ atmosphere in the context of a higher surface partial pressure and a climate that allowed liquid water on the planet's surface, the so-called ``warm and wet'' early Mars hypothesis, is one of the great riddles in Solar System science \citep{depater01,jakosky01,hurowitz17,diachille10,lammer13,lammer18,Palumbo2020,Scherf2021}.
Whether early Mars passed through a phase of a warm and wet climate with a much denser atmosphere during the Noachian eon or was predominantly cold and dry with sporadic warm phases is debated. Recently, several studies suggest that Mars most likely had a dense atmosphere during the Noachian eon, with an approximate surface pressure of at least 0.5\,bar \citep{hu15,jakosky18,kurokawa18}. \citet{kurokawa18} tried to reproduce the present atmospheric $^{14}$N/$^{15}$N and $^{36}$Ar/$^{38}$Ar isotope ratios by modeling atmospheric escape and volcanic outgassing and concluded that the present isotope ratios can only be reproduced if the atmospheric pressure had at least a value of $\sim$0.5 bar 4 billion years ago. The maximum atmospheric surface pressure at $\sim$3.6\,Ga was estimated from the size distribution of ancient craters by \citet{kite14} based on Mars Reconnaissance Orbiter high-resolution images. These researchers obtained an upper surface partial pressure limit of 0.9$\pm$0.1 bar, or 1.9$\pm$0.2 bar, if they excluded rimmed circular mesas by interpreting them as erosion-resistant fills or floors of impact craters. {The result by \citet{kite14} of 0.9\,bar was recently corrected with the same method by \citet{Warren2019} to be 1.1\,bar of maximum pressure.}

\citet{amerstorfer17} applied a sophisticated Monte Carlo model using the atmospheric and ionospheric profiles of \citet{tian09} corresponding to 1 (present), 3 (2.3\,-\,2.6\,Ga), and 10 (3.5\,-\,3.8 Ga), times the present solar EUV flux (EUV$_{\odot}$) to investigate the escape of suprathermal O and C atoms from the Martian upper atmosphere. These authors studied and discussed different sources of suprathermal O and C atoms in the thermosphere and their effects related to the varying EUV flux. Depending on the EUV activity of the young Sun, \citet{amerstorfer17} estimated that suprathermal atom escape from CO$_2$-dissociation products results in a CO$_2$ partial surface pressure between about $0.17-0.3$ bar at $\sim$\,4\,Ga, which is lower {than} the pressure inferred from the atmospheric $^{36}$Ar/$^{38}$Ar isotope reproduction attempts by \cite{kurokawa18}.

In particular at Venus and Mars, whose atmospheres are not protected by an intrinsic magnetic field, the interaction of the solar wind with the upper atmosphere plays an important role for atmospheric erosion over Solar System time scales \citep{luhmann91,luhmann92,fang13,lammer13,lammer18}. More recently, \citet{dong18} modeled the escape of O$^+$, O$_2^+$ and CO$_2^+$ by applying a 3D Mars global ionosphere thermosphere model of \citet{bougher15} that includes all the relevant neutral-ion chemistry and the radiative processes together with a 3D Mars multifluid magnetohydrodynamic solar interaction model \citep{najib11,dong14}. Similar as in \citet{tian09} and \citet{amerstorfer17}, atmospheric escape rates increase when the solar EUV flux exceeds $\sim 3$ EUV$_{\odot}$. The suprathermal O atom escape rates obtained by \citet{dong18} increase from $\sim 10^{25}$\,s$^{-1}$ (1 EUV$_{\odot}$) to $\sim 10^{26}$\,s$^{-1}$ (10 EUV$_{\odot}$) and are similar to those reported by \citet{amerstorfer17}. During the same time interval, the O$^+$ ion escape rate changed from $\sim 10^{24}$\,s$^{-1}$ to about $10^{27}$\,s$^{-1}$, indicating that atmospheric loss for early Mars is primarily controlled by ion escape \citep{dong18}. The main reason for this high ion escape rate is the expansion of the upper atmosphere when exposed to fluxes larger than 3 EUV$_{\odot}$.

Besides O$^+$ ion escape, \citet{dong18} also modeled the loss of O$_2^+$ and CO$_2^+$ for a radiation of 10 EUV$_{\odot}$ and obtained escape rates of $\sim$\,10$^{25}$\, s$^{-1}$ and $\sim 4\times 10^{24}$\,s$^{-1}$, respectively. These relatively low values are mainly due to the dissociation of these ions by the high EUV flux, leading to a low abundance of the molecules in the upper atmosphere and an increased escape of its dissociation products \citep{tian09}. From the $\sim$100 times higher O$^+$ ion escape rates at $\sim$4 Ga, \citet{dong18} concluded that their results are consistent with Mars having lost much of its atmosphere early in its history, causing the Martian climate to change from a warm and wet environment in the past to its present state. However, this conclusion might be questionable, since this particular study did neither include the escape of carbon-related dissociation products from CO$_2$ nor were the escape rates compared against realistic volcanic outgassing rates.

For better understanding the total atmospheric escape rates of carbon and oxygen related species during the Noachian eon, we invoke a 3D global hybrid model \citep{kallio03,dyadechkin13} {for the solar wind interaction with Mars} using the cold and hot atmospheric profiles for 1, 3 and 10 EUV$_{\odot}$ given in \cite{amerstorfer17}. Based on the O$^+$ and C$^+$ ions produced by photoionization of cold and suprathermal O and C atoms, we calculate their escape rates as well as the precipitation of these ions into the upper atmosphere. Upon entering these extended upper atmospheres, they can knock out other atmospheric species by transferring sufficient energy to even allow sputtered particles to leave the planet, {while a fraction can also contribute to the density of the hot population.}

We investigate the sputter efficiency not only for the main atmospheric species, but also for $^{36}$Ar and $^{38}$Ar noble gas isotopes which will then be used similar as in \citet{kurokawa18} to constrain the atmospheric CO$_2$ partial surface pressure evolution. Finally, we compare the total atmospheric CO$_2$ loss rates weighted by the escape of C within realistic volcanic CO$_2$ outgassing rates.

Section \ref{sec.Hot} contains a brief overview of how the density of hot particles, produced by chemical reactions in the upper atmosphere, are obtained. In Section \ref{sec.Hybrid}, the 3D global hybrid solar wind interaction model used to simulate the {escape and} precipitation of atmospheric ions is described, while Section \ref{sec.Yield} resumes the sputter model. Section \ref{sec.Volcanic} deals with the volcanic outgassing of Ar and in Section \ref{sec.Results} the results of the simulation are discussed; Section \ref{sec.Conclusion} concludes the work.

\section{Bulk atmosphere and hot O and C corona}\label{sec.Hot}
The background bulk CO$_2$-dominated atmosphere in the simulation for three different solar EUV fluxes, namely 1, 3, and 10\,EUV$_{\odot}$, consists of O, CO, CO$_2$, C, and O$_2^+$, CO$_2^+$, CO$^+$, and O$^+$ and was adopted from \citet{tian09} {and corresponds to a solar activity being in-between minimum and maximum activity}. The production of suprathermal atoms by various reactions in the upper atmosphere, their motion through the atmosphere up to the exobase {(i.e., 220\,km for 1\,EUV$_{\odot}$, 370\,km for 3\,EUV$_{\odot}$, and 750\,km for 10\,EUV$_{\odot}$)} and their density profiles above the exobase are simulated by means of a Monte Carlo model. Since this model is presented in detail in \citet{groeller10,groeller12,groeller14} and \citet{amerstorfer17}, we will only briefly describe the essential features.

\begin{table}[b!]
\caption{Sources of suprathermal O and C atoms.}
\label{tab:sources}
\centering
 \begin{tabular}{l l}
  \hline \\[-0.5cm]
  Source reaction hot O & Source reaction hot C \\[0.1cm]
  \hline \\[-0.3cm]
  \multicolumn{2}{c}{Dissociative recombination}  \\[0.2cm]
     CO$^+$ + e $\to$ C + O & CO$^+$ + e $\to$ O + C \\[0.2cm]
   CO$_2^+$ + e $\to$ CO + O &  CO$_2^+$ + e $\to$ O$_2$ + C\\[0.2cm]
  O$_2^+$ + e $\to$ O + O &\\[0.2cm]
  \multicolumn{2}{c}{Chemical reaction}  \\[0.2cm]
  O$_2^+$ + C $\to$ CO$^+$ + O & \\[0.2cm]
  \multicolumn{2}{c}{Photodissociation}\\[0.2cm]
  CO + $h\nu$ $\to$ C + O & CO + $h\nu$ $\to$ O + C \\[0.2cm]
  \hline
 \end{tabular}
\end{table}

We start with a number of source reactions, shown in Table \ref{tab:sources}, which produce the majority of suprathermal O and C in the upper atmosphere of Mars. At discrete altitudes, we determine the velocity distribution of the reaction products for a specific reaction and follow the newly born suprathermal atoms along their path through the thermosphere up to the exobase in the gravitational field of Mars. Through collisions with the background atmosphere, the suprathermal particles lose part of their initial energy on their way, while the collision partners from the background gain energy through such collisions on average and thus may become suprathermal. Total and differential cross sections are used to determine the collision probability and the energy transfer. At the exobase altitude, the energy distribution function of the suprathermal atoms is taken to calculate the exosphere density and the loss rates.

The solar flux for the photodissociation reactions is taken from SUMER/SOHO observations \citep{curdt01,curdt04}, where we have chosen observations from April 20, 1997, for quiet Sun conditions. The data were transferred to the orbit of Mars by dividing the photon flux by the square of the Sun-Mars distance. The atmospheric input profiles are assumed to represent average dayside conditions, thus a solar zenith angle of 60$^\circ$ was taken. To determine the solar flux for the considered solar zenith angle, we use the Chapman function for an isothermal atmosphere. The photodissociation and absorption cross sections are taken from the PHoto Ionization/Dissociation RATES database provided by \citet{huebner92}.

The kinetic energy transferred to the reaction products due to the exothermic source reaction is randomly chosen from the energy distribution of the corresponding reaction. The total kinetic energy in the center of mass frame, $E_{\rm{tot}} = E_{\rm{cm}} + E_{\rm{br}} + E_{\rm{v}} + E_r$, is used to determine the energy distribution. {For dissociative recombination (DR),} $E_{\rm{cm}}$ denotes the energy according to the relative velocity of the ion and the electron in the center of mass frame, and $E_{\rm{br}}$ is the released energy corresponding to the reaction channel. The vibrational and rotational energies are given by $E_{\rm{v}}$ and $E_r$, respectively. All molecules and atoms are assumed to be in their vibrational and rotational ground states. The total kinetic energy, $E_{\rm{tot}}$, is then distributed among the reaction products according to their masses. The components of the ion and electron velocity are taken randomly from a 1D Maxwell-Boltzmann distribution according to the temperature of the ions and the electrons. {A similar approach, as described here for DR, was also used for the other reactions listed in Table~\ref{tab:sources}.} More details about the computation of the energies are given in \citet{groeller10}.

The collisions between hot particles and the neutral background atmosphere can be elastic, inelastic, or quenching. While in elastic collisions the kinetic energy is conserved, in inelastic collisions part of the kinetic energy can be transferred into internal energy, i.e. vibrational energy or electronic excitation. An excited reactant will be de-excited and internal energy will be converted into kinetic energy during quenching collisions. For the treatment of the collisions and the corresponding total and differential cross sections, we adopt the same approach as \citet{groeller14}.

\begin{figure}[b]
\centering
\includegraphics[width=8cm,angle=0]{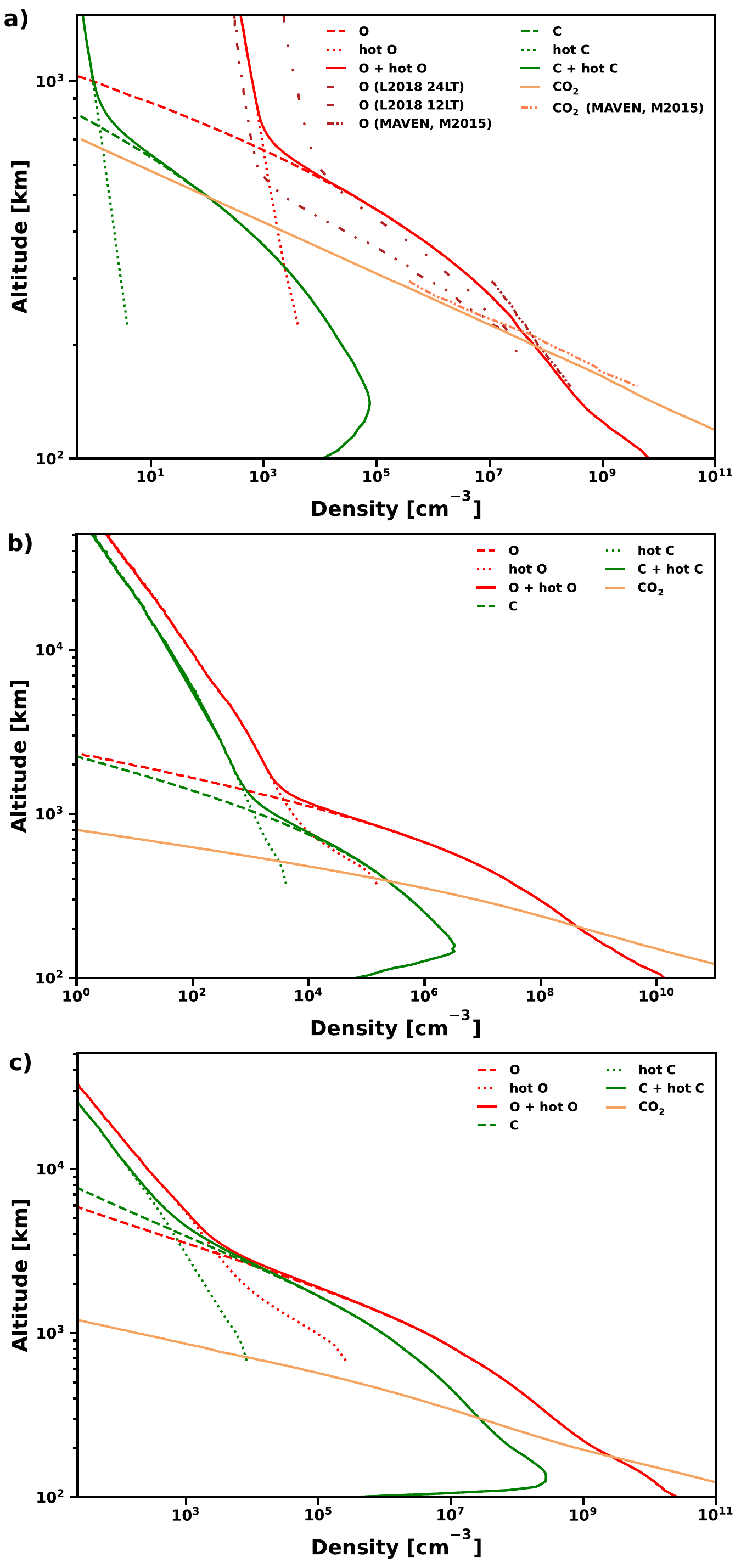}
\caption{Neutral density profiles of CO$_2$ molecules and C and O-related dissociation products {used in our simulations for present-day (a) and solar EUV fluxes that are 3 times (b) and 10 times (c) enhanced from now. For present-day, we also included MAVEN measurements for O and CO$_2$, as well as two O profiles from \citet{leblanc18} for comparison.}}
\label{Fig:Profiles}
\end{figure}\vspace{1mm}

 When hot particles arrive at the exobase, their motion above the exobase is assumed to be collisionless and to be described by ballistic orbits; {a fraction of these particles can be ionized, thereby forming an additional flux of precipitating pick-up ions}. The neutral density profiles for cold \citep{tian09,zhao15}, and hot C and O atoms modeled as described above are shown {for all EUV flux cases} in Fig.~\ref{Fig:Profiles}. It is seen that suprathermal C and O atoms produce highly extended hot atom coronae and are the dominant species for both cases. However, for a 10 times higher EUV flux, the upper atmosphere expands to great distances, because a larger amount of the infrared cooling molecule CO$_2$ is dissociated, which results in more heating and hence thermospheric expansion. As can be seen from Fig.~\ref{Fig:Profiles}, CO$_2$ molecules that are not dissociated remain close to the planet for both cases. In the following section we discuss the global hybrid model for the escape {and precipitation} of C$^+$ and O$^+$ ions which are dissociation products of CO$_2$ molecules from these extended upper atmospheres.

\section{Ion escape {and precipitation} modeling}\label{sec.Hybrid}
For the simulation of the solar wind with the Martian atmosphere, a global hybrid model, i.e. a semi-kinetic model, which treats the ions as particles and the electron as a massless fluid, is applied by using the upper atmosphere profiles shown in Fig.~\ref{Fig:Profiles}. Since a detailed description of the hybrid model can be found in \citet{kallio03} and \citet{dyadechkin13}, only a brief overview of the model will be given in the following.

\subsection{Basic equations of the hybrid model}

The hybrid model includes the following equations:

\begin{eqnarray}
 \nabla\times{\bf B} &=& \mu_0{\bf j} \label{Ampere_law} \\
 {n_e} &=& \mid{e}\mid^{-1}\sum_i q_in_i  \label{qn} \\
 {\bf j} &=& \sum_i q_i n_i {\bf v}_i + en_e{\bf U}_e \label{tot_currnent} \\
 {\bf E} &=& -{\bf U}_e\times{\bf B} + \frac{\nabla{p}_e}{en_e} + \frac{{\bf j}}{\sigma} \label{Ohm} \\
 \frac{\partial{\bf B}}{dt} &=& - \nabla\times{\bf E} \label{faraday_law} \\
 \frac{d{\bf v}_i}{dt} &=& \frac{q_i}{m_i}\biggr({\bf E} + {\bf v}_i\times{\bf B}\biggr) \label{Lorentz} \\
 \frac{d{\bf x}_i}{dt} &=& {\bf v}_i \label{x_propagate}
\end{eqnarray}

Equation~(\ref{Ampere_law}) represents Amp$\grave{e}$re's law, where ${\bf B}$ is the magnetic field, ${\bf j}$ is the total electric current density and $\mu_0$ is the magnetic permittivity in a vacuum. Equation~(\ref{qn}) states the quasi-neutrality of electric charge, where $n_e$, $e$, $n_i$ and $q_i$ are the number density and charge of the electron and ions, respectively. Equation~($\ref{tot_currnent}$) is the total electric current density
, i.e. the sum of the electric current of electrons and ions, with ${\bf v}_i$ and ${\bf U}_e$ being the velocity of the ions and the bulk velocity of the electron fluid, respectively. Equation~($\ref{Ohm}$) is generalized Ohm's law, where $\nabla{p}_e$ is the gradient of the electron pressure (assuming an isothermal electron fluid with $p_e = n_e k_B T_e$, where $T_e = 10^5$ K) and $\sigma$ the electric conductivity. Equation~($\ref{faraday_law}$) is Faraday's law, while Equation~($\ref{Lorentz}$) represents Newton’s law of motion, including the Lorentz force and with $m_i$ being the mass of the ions. Finally, the last Equation~($\ref{x_propagate}$) defines the velocity of the ions. {In the simulations, the ionospheric obstacle to the solar wind flow is modeled as a superconducting sphere. This is implemented by setting the resistivity at zero inside the inner boundary and the magnetic field does not diffuse through the obstacle. Further, ions impacting the inner boundary are removed from the hybrid model and are then used as sputter agents at the exobase level in the sputtering model.}

\subsection{Simulation domain and input parameters}
The following right handed cartesian coordinate system was used in the simulations: The $x$-axis points from the center of the planet against the solar wind flow which is assumed to flow along the Mars-Sun line, the IMF component perpendicular to the solar wind flow is along the $y$-axis and, thus, the convection electric field is along the $z$-axis. The size of the simulation box extends from -15\,$R_{\rm M}$ to 15\,$R_{\rm M}$, with $R_{\rm M}$ as the radius of Mars ($R_\textrm{M}$ = 3390 km)
, in all three coordinate directions with $123\times123\times123$ grid cells, each having a size of $847.5\,\rm{km} = R_{\rm M}/4$. Mars is assumed to be unmagnetized, the solar wind consists of a single species H$^+$ and the simulation timestep is 50\,ms.

The three global hybrid simulations performed in this study used the same grid cell and simulation domain sizes to keep the model results comparable to each other. This means that the spatial resolution was quite coarse near the inner boundary (see Table~\ref{tab:Hybrid}), which can affect the morpohology of the magnetic and electric field at low altitudes. {However, the oxygen ion escape and precipitation rates in the 1\,EUV$_{\odot}$ case were found to be of the same order of magnitude than in our earlier simulations for Mars under moderate solar EUV conditions suggesting that the used spatial resolution does not affect our conclusions  \citep{Brain2010,kallio10}.}

In the simulations, the planetary ions are obtained by photoionization {(see Table~\ref{tab:tbl1})} of four different atmospheric populations: cold and hot O, and cold and hot C (see Fig.~\ref{Fig:Profiles}). {The photoionization rates in the 1\,EUV$_{\odot}$ case are estimated as follows. The state distributions from DR are assumed for simplicity at the zero relative collision (detuning) energy given in \citet{Rosen1998} as C(1D): 14.5\% and C(3P): 76.1\% + 9.4\% for carbon, and O(1D): 9.4\% and O(3P): 76.1\% + 14.5\% for oxygen. The photo rate coefficients of different states for the Quiet Sun and Active Sun at 1\,AU are taken from the PHoto Ionization/Dissociation RATES database\footnote{\url{https://phidrates.space.swri.edu}, referenced on August 30, 2021} \citep{Huebner1979,huebner92,Huebner2015}. The total ionization rate is the weighted average of the photo rate coeffients, where the weight is the portion of the state distribution. Finally, the total rates in the 1\,EUV$_{\odot}$ run are an average between the quiet and active Sun and scaled from 1\,AU to Mars orbital distance. The photoionization rate coefficients in the 3\,EUV$_{\odot}$ and 10\,EUV$_{\odot}$ cases are the 1\,EUV$_{\odot}$ case values multiplied by the factors of 3 and 10, respectively. Table~\ref{tab:tbl1} lists the coefficients.}

{Hot particles are defined by having an energy $E > 1.5 \times E_{\rm therm}$  and cold particles by $E < 1.5 \times E_{\rm therm}$ with $E_{\rm therm}$ being the thermal energy (a detailed discussion on this distinction can be found in \citet{amerstorfer17}).} Since the cold population is only available up to the exobase, its density has been extrapolated by assuming a Maxwell distribution with the given temperature at the exobase. { The densities of the hot particles up to 15\,$R_{\rm M}$ are extrapolated by using their actual (non-Maxwellian) distribution function at the exobase.}

{As inputs for the three global hybrid simulations we fit the retrieved densities by the following functions, i.e.,}

\begin{description}
  \item[1 EUV$_{\odot}$:]
    \begin{eqnarray}
      \mathrm{Cold~O:}~n(h) &=& 1.3252\times10^{15}\exp(-1.5425\times10^{-5} h) \\
      \mathrm{Cold~C:}~n(h) &=& 1.9586\times10^{11}\exp(-1.0476\times10^{-5} h) \\
      \mathrm{Hot~O:}~n(h) &=& 3.0158\times10^{17} h^{-1.4434} \\
      \mathrm{Hot~C:}~n(h) &=& 2.0487\times10^{14} h^{-1.401}
    \end{eqnarray}
  \item[3 EUV$_{\odot}$:]
    \begin{eqnarray}
      \mathrm{Cold~O:}~n(h) &=& 2.0\times10^{78} h^{-11.4} \\
      \mathrm{Cold~C:}~n(h) &=& 1.0\times10^{65} h^{-9.406} \\
      \mathrm{Hot~O:}~n(h) &=& 1.2104\times10^{20} h^{-1.7454} \\
      \mathrm{Hot~C:}~n(h) &=& 6.1269\times10^{16} h^{-1.3262}
    \end{eqnarray}
  \item[10 EUV$_{\odot}$:]
    \begin{eqnarray}
      \mathrm{Cold~O:}~n(h) &=& 1.82\times10^{54} h^{-6.92} \\
      \mathrm{Cold~C:}~n(h) &=& 1.55\times10^{44} h^{-5.35} \\
      \mathrm{Hot~O:}~n(h) &=& 1.1638\times10^{26} h^{-2.5015} \\
      \mathrm{Hot~C:}~n(h) &=& 4.781\times10^{22} h^{-2.0841}
    \end{eqnarray}
\end{description}

{Here, $h$ is the altitude above the surface in m, and $n(h)$ is the particle density in $m^{-3}$.}

\begin{table}
\caption{{Photoionization rates of oxygen and carbon ions used in the global hybrid simulation for Mars at 1.5\,AU. }}
\label{tab:tbl1}
\centering
\begin{tabular}{l l l}
Run & O$^+$ [s$^{-1}$] & C$^+$ [s$^{-1}$]\\
\hline
\hline\\
EUV 1  & {$1.944 \times 10^{-7}$} & {$8.571 \times 10^{-7}$}\\
EUV 3  & {$5.831\times 10^{-7}$} & {$2.571 \times 10^{-6}$}\\
EUV 10 & {$1.944 \times 10^{-6}$} & {$8.571 \times 10^{-6}$}\\
\hline
\end{tabular}
\end{table}

Young solar-like stars show a wide variety of different rotation rates which strongly influences their mass loss and EUV flux evolution \citep{tu15,johnstone15a,johnstone15b}. Even though solar-like stars start with similarly high EUV flux and mass loss rates, the irradiation and stellar winds of slow rotators decline much faster than for moderate or even fast rotators. After about one billion years, however, all stars converge again towards one common rotational track. For our simulations, we assumed the Sun to be born as a slow rotator which is in agreement with several recent studies \citep[e.g.][]{saxena19,lammer20,Johnstone2021}. Therefore, our model runs for 3 and 10 EUV$_{\odot}$ correspond to about 2.6 and 3.8\,Ga, respectively, \citep{tu15,amerstorfer17}.

For estimating the solar wind parameters (see Table~\ref{tab:Hybrid}), i.e. the interplanetary magnetic field $B_{\rm IMF}$, solar wind temperature $T_{\rm sw}$, velocity  $V_{\rm sw}$ and density $n_{\rm sw}$ , we use the 1D thermal pressure driven hydrodynamic stellar wind evolution model for low-mass main-sequence stars developed by \citet{johnstone15b}. The model is calibrated with present-day solar wind data and assumes that the mass loss rate $\dot{M}_\star$ depends on the rotation rate $\Omega_\star$ as $\dot{M}_\star \propto \Omega_\star^a$ with $a \sim 1.33$ \citep[Model~A of][]{johnstone15a} while the wind temperature scales linearly with the coronal temperature.

Even though we aligned our model runs to the evolution of a slow rotator, the results can be easily adapted for moderate and fast rotators, since the only parameter that changes significantly is the time in the past for which the EUV flux of the Sun coincides with 3 and 10 EUV$_{\odot}$. The solar wind parameter themselves, however, vary insignificantly for the same EUV fluxes within different rotational tracks.

\begin{table}[h!]
\caption{Simulation parameters}\vspace{2mm}
\label{tab:Hybrid}
\centering
\resizebox{\textwidth}{!}{%
 \begin{tabular}{|c|l|l|l|l|l|l|l|l|l|}\hline
  EUV level & $B_x [nT]$ & $B_y [nT]$ & $B_z [nT]$ & $n_{\rm sw}$ [cm$^{-3}$] & $V_{\rm sw}$ [km/s] & $T_{\rm sw} [K]$ &  {$T_{\rm exo} [K]$} & exobase altitude [km] & {hybrid model} \\
& & & & & & &  & & {inner boundary} [km] \\\hline
 1 & 0 & 1.08 & 0 & 1.83 & 475.8 & $5.5\times 10^4$ & {222} & {220} & 300 \\
 3 & 0 & 1.81 & 0 & 2.54 & 574.8 & $7.3\times 10^4$ & {416} & {370} & 370 \\
10 & 0 & 2.93 & 0 & 3.42 & 693.9 & $9.7\times 10^4$ & {806} & {750} & 850 \\\hline
 \end{tabular}}
\end{table}


\section{Atmospheric sputtering}\label{sec.Yield}
The estimation of the atmospheric sputter yields caused by the precipitating pick-up O$^+$ and C$^+$ ions is based on the model of \citet{johnson90,johnson94}. The sputter yield $Y_i$, i.e. the number of species $i$ ejected per incident ion, basically consists of two contributions: the ejection of particles by a single collision with the incoming ion, $Y^s$, and the ejection due to the cascade of collisions initiated by the incoming ion, $Y^c$, approximated by \citep{johnson90}
\begin{equation}\label{eq:yield1}
Y^s\approx\bar{P}_{es}^i\frac{\sigma(T>U_{0i})}{\sigma_d(E_{es})\cos\theta},\hspace{5mm}
 Y^c\approx\frac{\beta}{2}\frac{\alpha\, S_n(E_{in})}{U_{0i}\sigma_d(E_{es})}\frac{1}{(\cos\theta)^p}.
\end{equation}
Here, $\bar{P}_{es}^i$ is the escape probability for species $i$, $\sigma(T>U_{0i})\equiv\int_{U_{0i}}d\sigma$ is the collision cross section for a particle receiving an energy transfer $T$ sufficiently high to enable escape, where $U_{0i}$ is the gravitational binding energy of species $i$ at the exobase, $\sigma_d(E_{es})$ is the momentum cross section of species $i$ escaping with energy $E_{es}$, and $\theta$ is the polar angle of the incident ion. Further, $S_n$ is the nuclear stopping cross section, expressed by \citep{johnson90},
\begin{equation}
S_n=\frac{\gamma E_{in}}{2}\sigma_d(E_{in}),\hspace{5mm}\gamma=\frac{4m_{in}m_{es}}{(m_{in}+m_{es})^2},
\end{equation}
where $m_{in}$ and $m_{es}$ are the masses of the incident and escaping particle, respectively, and where the parameter $\alpha$ depends on the mass ratio $m_{es}/m_{in}$,$\beta$ on the interaction potential, and $p$ mainly on the energy of the precipitating ions. Finally, the cross section for transferring an amount of energy between $E$ and $E+dE$ to an initially stationary atom can be written as \citep{sieveka84}
\begin{equation}
\frac{d\sigma}{dE}=\frac{\varepsilon^2}{E_{max}}\frac{d\sigma}{dt},
\end{equation}
with the scaled energy transfer $t$ given by \citep{johnson90},
\begin{equation}
t=\varepsilon^2\frac{E}{E_{max}},\hspace{5mm} E_{max}=\gamma E_{in},\hspace{5mm} \varepsilon=\frac{\gamma E_{in}a_u}{2A},
\hspace{5mm} A=\frac{2m_{in}}{m_{in}+m_{es}}Z_{in}Z_{es}e^2,
\end{equation}
where $a_u$ is the screening radius, $Z_{in}$ and $Z_{es}$ are the nuclear charge of the incident and escaping particle, respectively, and $e$ the electron charge.

Using Gaussian units and a universal interaction potential, the screening radius and the nuclear stopping cross section can be approximated via \citep{johnson90}
\begin{equation}
a_u=\frac{0.8853\, a_0}{Z_{in}^{0.23}+Z_{es}^{0.23}},\hspace{5mm}
 S_n=2\pi\frac{A^2}{\gamma E_{in}}[2\varepsilon s_n(\varepsilon)],
\end{equation}
where
\begin{equation}
[2\varepsilon s_n(\varepsilon)]=\left\{
\begin{array}{ll}
\ln\varepsilon &\varepsilon>30\\
\ln(1+1.138\,\varepsilon)/(1+0.0132\,\varepsilon^{-0.787}+0.196\,\varepsilon^{0.5})&\varepsilon<30\\
\end{array}
\right.
\end{equation}
and with
\begin{equation}
\frac{d\sigma}{dt}=\frac{\pi a_u^2}{2}\frac{f(t^{1/2})}{t^{3/2}},\hspace{5mm} f(\varepsilon)=
 \frac{d}{d\varepsilon}[\varepsilon s_n(\varepsilon)].
\end{equation}
Using 'standard' values for the parameters $\alpha,\beta$ and $p$ \citep{johnson94}, the total sputter yield of species $i$ can then be written as
\begin{equation}
Y_i=c_i(Y^s+Y^c)=
\frac{c_i}{\sigma_d(\bar{E}_{es})}\left[\frac{0.5\,\sigma_A(T>U_{0i})}{\cos\theta}+
\frac{3}{\pi^2}\frac{\alpha S_n(E_{in})}{(\cos\theta)^{1/6}U_{0i}}\right],
\end{equation}
with $c_i$ the concentration of species $i$ at the exobase and $\bar{E}_{es}$ the escape energy averaged over the distribution of escape energies, where the latter is approximated by (\cite{sieveka84})
\begin{equation}\label{eq:yield-end}
f(E)\propto\frac{1}{E^2}\left(1-\sqrt{\frac{E}{\gamma E_{in}}}\right).
\end{equation}

Although sputtering can potentially eject any atmospheric species residing at the exobase, it is especially interesting to investigate its influence on $^{36}$Ar and $^{38}$Ar isotopes. Argon {has one of the highest ionization thresholds \citep[e.g.,][]{Zahnle2019} and}, similar as CO$_2$, a high escape energy, thus a mass-related escape of argon isotopes and hence its fractionation can only be caused by sputtering from precipitating pick-up ions \citep{jakosky17}. Consequently, the escape of the atmospheric argon content on Mars after cessation of its magnetic dynamo should solely depend on sputtering. Replicating the loss of argon over time would therefore allow us to constrain possible pathways for the atmospheric development of Mars.

The Martian meteorite ALH84001 with an age of $\sim 4.1$ billion years \citep{lapen10} encloses isotopic imprints of the ancient Martian atmosphere.
Its $^{36}$Ar/$^{38}$Ar ratio of $\sim 5.3$ \citep{mathew01} is higher than the current  Martian atmospheric value of $4.2\pm 0.1$ \citep{atreya13} {but similar to the value of the Earth's present-day atmosphere of $^{36}$Ar/$^{38}$Ar\,=\,$5.32 \pm 0.33$ \citep{marty12}. This decrease of the ratio over time} suggests that sputtering preferentially removed the lighter $^{36}$Ar isotope from the upper atmosphere. The reason for this preference comes from the diffusive separation of species occurring between the homopause and exobase \citep{wallis89,bauer04,lammer13}. Above the homopause level, turbulent mixing becomes increasingly less efficient until gravity dominates and diffusion sets in. Hence, lighter atoms and molecules become more abundant with height than heavier ones. Since escape occurs at the exobase, lighter particles are therefore preferably removed. The ratio $R$ of the abundances of two species at the exobase to that at the homopause can be written as \citep{jakosky94}
\begin{equation}
R=\exp\left(-\frac{{\Delta}m g\Delta{z}}{k T}\right),
\end{equation}
where ${\Delta} m$ is the difference in mass between a lighter and a heavier atom or molecule (e.g. CO$_2$ and $^{36}$Ar, $^{38}$Ar), ${\Delta} z$ is the difference between the homopause and the exobase altitude, $g$ is the gravitational acceleration, and $T$ is the average temperature at this distance.

To estimate the abundance of both $^{36}$Ar and $^{38}$Ar isotopes at the exobase level we apply this approach to the atmospheric profiles shown in Fig.~\ref{Fig:Profiles} and discussed in Section~\ref{sec.Hot}. As these argon isotopes make up only a small fraction of the Martian atmosphere and sputtering is, to first order, only dependent on the relative concentration at the exobase and the necessary escape energy of the species, the sputter loss can be scaled to the loss of the dominant species CO$_2$ \citep{johnson00}.

\citet{hutchins96} proposed to calculate the evolutionary loss $L^i_{\rm sp}$ of $^i$Ar through sputtering via the relation
\begin{equation}\label{Eq:Lsput}
L^i_\mathrm{sp} = [{\rm CO_2}]_\mathrm{sp}\frac{Y({\rm ^{i}Ar_{hp}})}{Y(\mathrm {CO_{2,hp}})}\left[\frac{^i\mathrm{Ar}}{\mathrm{CO_2}}\right]_{\rm hp}R_{\rm ^iAr/CO_2}\,s_{\rm f},
\end{equation}
with [CO$_2]_\mathrm{sp}$ being the amount of CO$_2$ sputtered per time unit, $Y(^i\mathrm{Ar})/Y(\mathrm{CO}_2)$ represents the ratio of the sputtering yield and $[^i\mathrm{Ar}/\mathrm{CO}_2]_{hp}$ is the mixing ratio of $^i$Ar at the homopause, with present day values taken from MAVEN measurements \citep{jakosky15}; the factor $s_{\rm f}$ allows to take into account possible uncertainties that are related to the sputter escape estimates. {CO$_{2,\mathrm{hp}}$} is calculated from the partial CO$_2$ surface pressure
$P_{\rm CO_2}$ at each time-step \citep{leblanc12,slipski16,kurokawa18} with the simple relation
\begin{equation}\label{eq:co2}
\mathrm{CO}_{2,\rm hp} = \frac{4\pi P_{\rm CO_2} R_{\rm M}^2}{m_{\rm CO_2}g},
\end{equation}

where $m_{\rm CO_2}$ is the mass of CO$_2$. This approach is valid since 99\,\% of the atmospheric mass is located in the homosphere and the abundance in the upper atmosphere has only a negligible effect on the surface pressure. Additionally, Equation~(\ref{Eq:Lsput}) actually only requires the ratio of the argon isotopes to CO$_2$ at the homopause, which due to efficient mixing is equivalent to the ratio found in the entire homosphere.

{We will calculate argon sputtering through both methods, i.e. through the simple scaling with CO$_2$ sputtering and CO$_2$ pressure (Equation~\ref{Eq:Lsput}) and directly through calculating how many argon particles are sputtered by the precipitating C$^+$ and O$^+$ ions as retrieved from our hybrid simulations. For both calculation methods, we use the yields, as estimated with Equations (\ref{eq:yield1}\,--\,\ref{eq:yield-end}). The first method has the advantage that it can give us an initial CO$_2$ pressure range for which the $^{36}$Ar/$^{38}$Ar ratio at 4.1\,Ga can be reproduced, but with argon sputtering being simply scaled to CO$_2$ sputtering and not directly estimated from the precipitating ions that are retrieved from our hybrid simulations. The second method, on the other hand, calculates argon sputtering directly, but cannot give a range for the initial pressure.}

\section{Volcanic outgassing of argon isotopes and crustal production}\label{sec.Volcanic}
Volcanic outgassing has been an important source of atmospheric constituents over most of Mars` history \citep{jakosky01,manning06,grott11,grott13}. Here we assume, as in \citet{morschhauser11}, that part of the mantle undergoes partial melting and produces a new crust on the surface, while releasing volatiles into the atmosphere. We use the following relation from \citet{leblanc12} and \citet{slipski16} for the estimation of the outgassed Ar isotopes,
\begin{equation}
S^i_{\rm out}(t) = \frac{^{i}Ar_{\rm m} F_{\rm c}(t)}{V_{\rm m}(t)}(t)v_{\rm f}.
\end{equation}
Here, {$F_{\rm c}$(t)} denotes the crustal production rate, i.e., how much volume of the mantle is added to the crust per time unit, {$V_{\rm m}$(t)} is the total mantle volume and $^{i}Ar_{\rm m}$ is the abundance of the argon isotope $i$ in the mantle. The additional factor $v_{\rm f}$ can be used to scale the efficiency of this process, since the volcanism rate, ratio of intrusive to extrusive volcanism, enrichment of the argon isotopes and other factors influencing the outgassing are not perfectly constrained. According to \citet{slipski16} and based on the {implemented} crustal production rates {$F_{\rm c}$} of \cite{greeley91} (later than $\sim 3.8$ Ga) and  \cite{morschhauser11} (earlier than $\sim 3.8$ Ga), the factor $v_{\rm f}$ lies within 0.3\,-\,1.0. The specific evolution of the argon isotope abundances can then be calculated via
\begin{equation}
\frac{d^{i}Ar}{dt}=S^i_{\rm out}(t) - L^i_{\rm sp}(t),
\end{equation}
where the source term $S^i_{\rm out}$ is the outgassed amount of $^i$Ar and $L^i_{\rm sp}$ is the sputter loss of $^i$Ar. The total change will be evaluated at each time-step, {starting from the time range when the Martian magnetic field is thought to have ceased between 3.6\,Ga and 4.1\,Ga \citep{milbury12,lillis13,Mittelholz2020} until the present-day.}

Since the $^{36}$Ar/$^{38}$Ar ratio of ALH 84001 of $\sim 5.3$ is comparable to that of carbonaceous chondrites \citep{marty12,mazor70}, no significant fractionation prior to its creation could have occurred. The lack of any fractionation earlier than $\sim 4.1$ Ga
might either be due to the Martian intrinsic magnetic field preventing fractionation through sputtering and/or due to a complete loss of the whole atmosphere prior to $\sim 4.1$ Ga and a subsequent replenishment through volcanic outgassing or impact delivery. The latter argument is supported by the isotopic ratio of $^{14}$N/$^{15}$N in ALH~84001, which shows no significant deviation from the chondritic value \citep{avice20}, while the present-day ratio is much {lower}, indicating non-thermal escape processes of nitrogen after the formation of the meteorite, such as dissociative recombination of N$_2^+$ as shown by \citet{Fox1997}. We note however, that, contrary to argon, nitrogen could have also been fractionated due to its lighter weight through ion escape or even thermal loss \citep[see e.g.][for a discussion on fractionation through different escape processes]{lammer20b} at the time when Mars still maintained its intrinsic magnetic field. The lack of any notable fractionation prior to 4.1\,Ga, therefore, supports the idea that Mars lost its entire atmosphere early on \citep{Scherf2021}.

Since our estimated sputtering rates are based on the modeled planetward-scattered exospheric ion flux, we also know the total related escape rate from C$^+$, CO$^+$, CO$_2^+$, O$^+$, etc. Because of this, we can then analyze scenarios for which the present day atmospheric $^{36}$Ar/$^{38}$Ar isotope ratio can be reproduced in dependence on the maximum possible loss of the planet's CO$_2$ atmosphere over time, including the role of volcanic outgassing during Mars' history.

\section{Results and Discussion}\label{sec.Results}

\begin{figure}[b]
\centering
\includegraphics[width=12cm,angle=0]{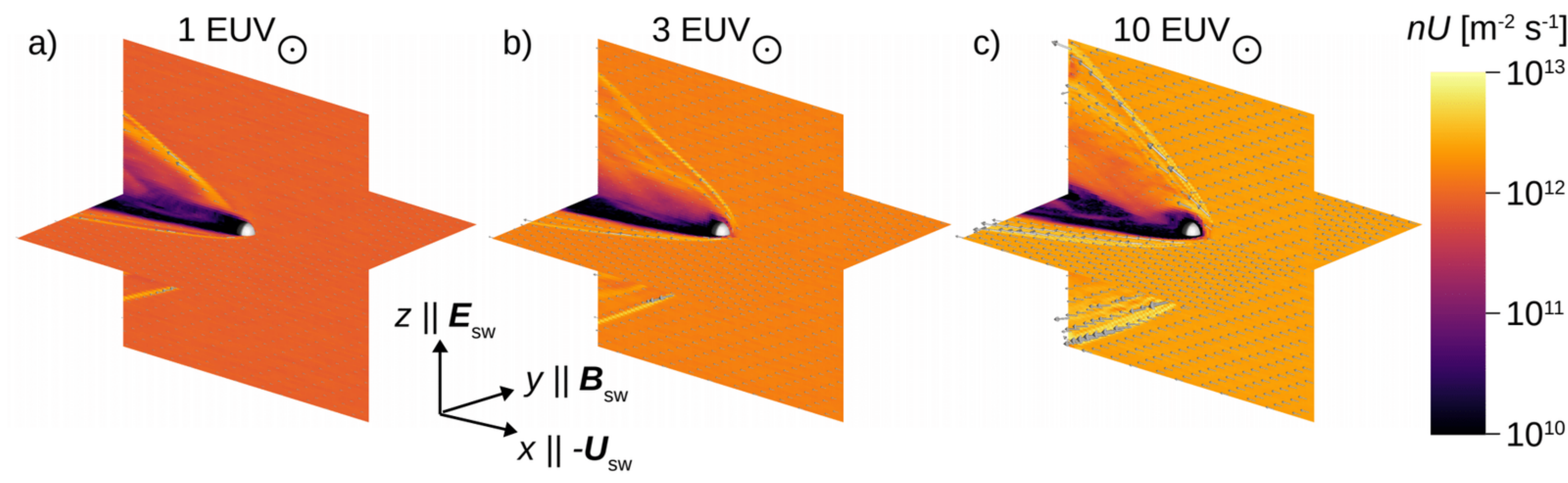}
\caption{An overview of the Martian induced magnetosphere in the three global hybrid simulation runs. The coloring on the $xy$ ($z=0$) and $xz$ ($y=0$)
planes gives the solar wind proton bulk number flux. Grey vectors show
the morphology of the solar wind bulk flow and the vectors have the same
scaling in each panel. Black arrows show the orientation of the
coordinate axes and upstream undisturbed solar wind convection electric
field ($\vec{E}_\textrm{sw}$), IMF ($\vec{B}_\textrm{sw}$) and bulk
velocity ($\vec{U}_\textrm{sw}$) vectors.
}
\label{Fig:Hybrid.SWFlux}
\end{figure}

The global hybrid simulation runs for the three EUV cases are illustrated in Fig.~\ref{Fig:Hybrid.SWFlux}. In all the runs, the induced magnetosphere is evident with the bow shock as the outermost boundary, where the flux increases suddenly. Downstream of the bow shock is the turbulent magnetosheath, which separates the induced magnetospheric boundary and the low density wake from the bow shock and the upstream solar wind. In the nominal 1 EUV$_{\odot}$ run, the induced magnetosphere is the most compact and the bow shock closest to the planet. The size of the induced magnetosphere and the bow shock distance increase in concert with the increasing EUV flux. In the 10 EUV$_{\odot}$ run, the induced magnetosphere is considerably larger compared to the 1 and 3 EUV$_{\odot}$ runs. The EUV increase is associated with an increase in the height of the model exobase and an enhanced planetary ion production, leading to, e.g., enhanced massloading of the solar wind flow.

\begin{figure}[h]
\centering
\includegraphics[width=10cm,angle=0]{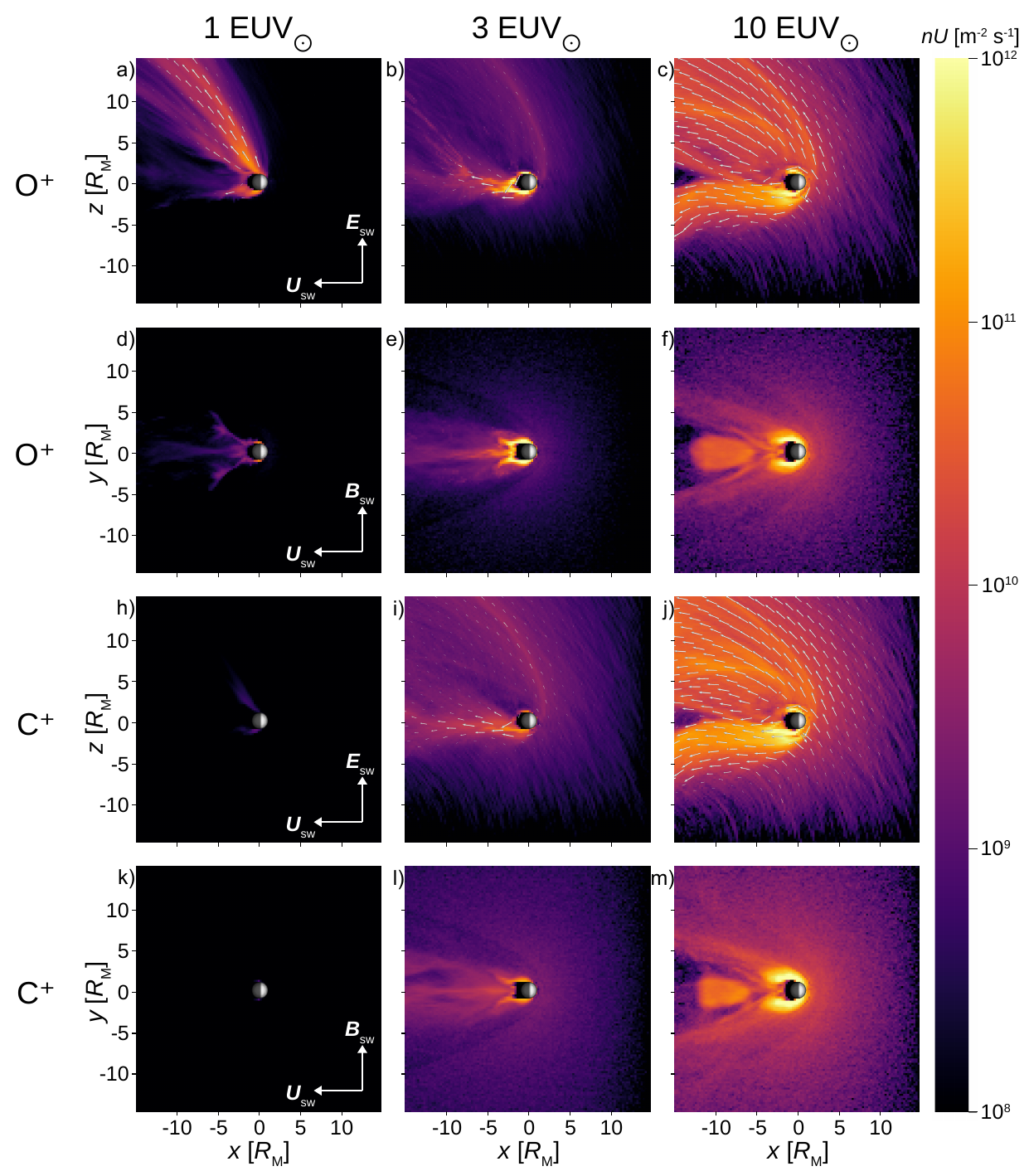}
\caption{Planetary ions in the three global hybrid simulation runs. The coloring
on the $xz$ ($y=0$) and $xy$ ($z=0$) planes gives the bulk number flux.
Grey vectors show the morphology of the planetary ion bulk flow per
species and the vectors have the same scaling in each panel. Two top
rows are the O$^+$ ions and the two bottom rows are the C$^+$ ions.}
\label{Fig:Hybrid.IonFlux}
\end{figure}

Fig.~\ref{Fig:Hybrid.IonFlux} illustrates the escaping planetary O$^+$ and C$^+$ ions from Mars in the global hybrid simulation. In all the runs, a notable asymmetry of planetary ions in the $xz$-plane between the $z>0$ and $z<0$ hemispheres is seen due to the solar wind pickup of heavy ions, whereas as the $xy$-plane (dawn-dusk plane) is symmetric due to the zero upstream flow-aligned IMF component $B_x$ \citep{jarvinen14}. In the 1 EUV$_{\odot}$ run, the planetary ions follow the morphology of the nominal Mars-solar wind interaction: there is the pickup ion plume along the convection electric field $E_{sw}$ and the tailward escape in the wake \citep{kallio10,dong17}. In the 3 and 10 EUV$_{\odot}$ runs, the system becomes heavily massloaded and the interaction appears more comet-like, with a dense ion cloud at low altitudes and pickup and tailward escape at higher altitudes, compared to the nominal run.

\subsection{Escape Rates}

Table~\ref{Tab:Hybrid.Rates} summarizes the production, impact and escape rates of the planetary ions obtained from the hybrid simulations for the three different EUV cases. Since the initial energy of newly born hot ions is higher than those of newly born cold ions, the former start with larger gyroradii and can therefore escape more easily than the cold ions with small gyroradii. This effect is most distinct for the 1 EUV$_{\odot}$ case, while for {3 and 10 EUV$_{\odot}$}, also most of the cold ions already have sufficient initial energies in order to escape.

{The ratio of escape rate to the production rate of planetary ions increases with increasing EUV in the global hybrid simulation runs. This results from the expansion of the neutral profiles with higher EUV conditions. When higher planetary neutral densities, and, thus, higher planetary ion production, occur at higher altitudes, less of those ions find their way back near the planet for precipitation. In other words, for higher EUV fluxes the planetary ions massload the solar wind at a higher rate relative to the production rate, as suggested in ealier studies \citep{Chaufrey2007}. This is especially clear for the cold population, whereas hot populations are quite extended already for the 1\,EUV$_{\odot}$ case and most of the produced hot ions escape in all runs. An extreme example of the higher massloading rate is the cold O$^+$ impact rates, which decrease from 3\,EUV$_{\odot}$ to 10\,EUV$_{\odot}$ \citep{Martinez2019}.}

\begin{figure}[h]
\centering
\includegraphics[width=12cm,angle=0]{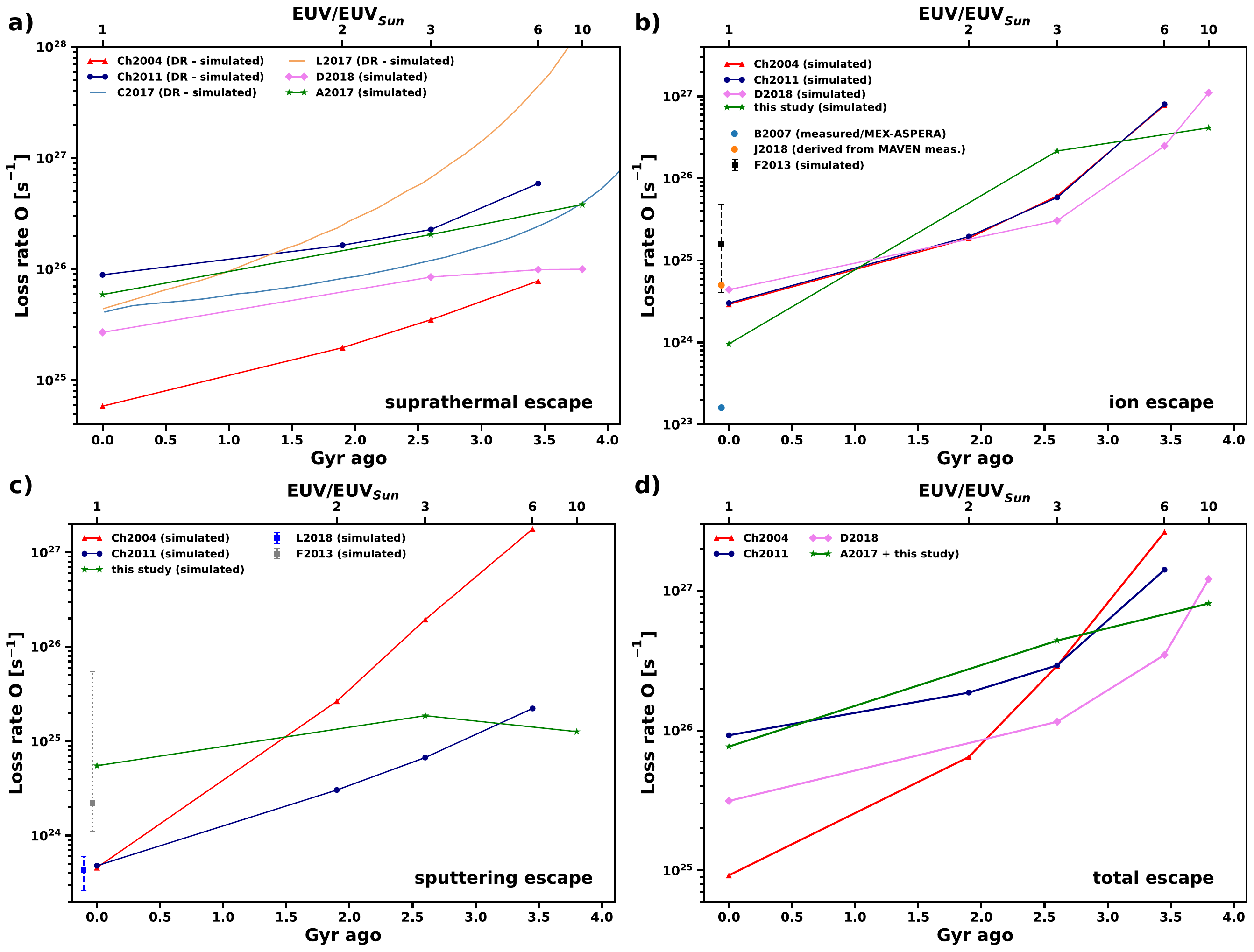}
\caption{Oxygen escape rates (a) dissociative escape; b) ion escape; c) sputtering; d) total escape of a-c) from our present and previous work A2017 \citep{amerstorfer17} compared with several other studies. Here, Ch2004 refers to \citet{chassefiere04}, Ch2011 to \citet{chassefiere11}, C2016 to \citet{cravens17}, L2017 to \citet{lillis17}, F2013 to \citet{feng13}, D2018 to \citep{dong18}, L2018 to \citet{leblanc18} {-- all of them simulations; B2007 refers to MEX/ASPERA measurements by \citet{Barabash2007}, and J2018 to \citet{jakosky18}, who derived ion escape values from MAVEN measurements.} The abbreviation DR stands for dissociative recombination.}
\label{Fig:EscapeRatesOxygen}
\end{figure}

\begin{figure}[h]
\centering
\includegraphics[width=12cm,angle=0]{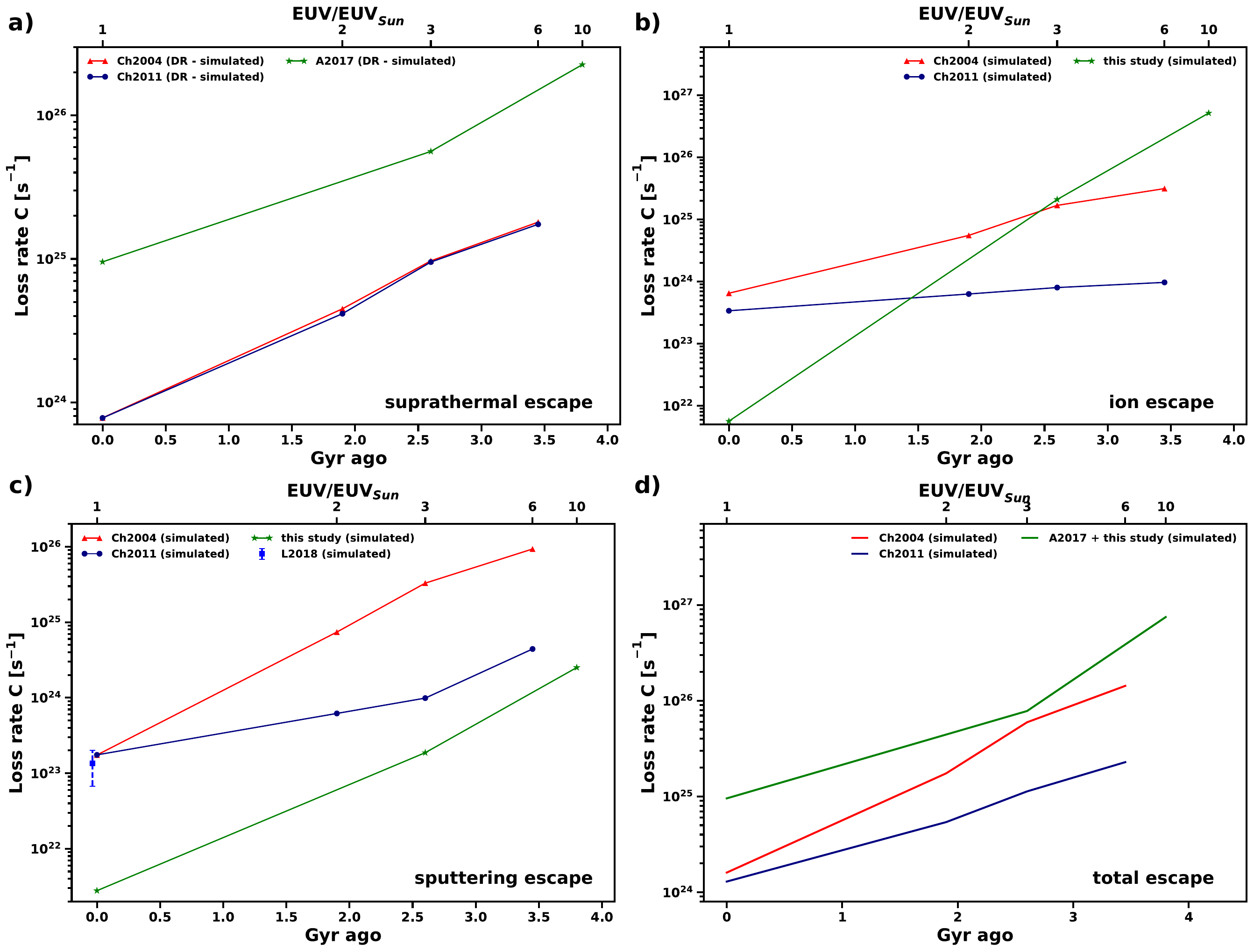}
\caption{Carbon escape rates (a) suprathermal; b) ion escape; c) sputtering; d) total escape of a-c) from our present and previous work A2017 \citep{amerstorfer17} compared with several other studies. Here, Ch2004 refers to \citet{chassefiere04}, Ch2011 to \citet{chassefiere11}, L2018 to \citet{leblanc18}. The abbreviation DR stands for dissociative recombination.}
\label{Fig:EscapeRatesCarbon}
\end{figure}

For present Mars (1 EUV$_{\odot}$), the total O$^+$ and C$^+$ escape rates are found to be {$\sim 9.6\cdot 10^{23}$ s$^{-1}$ and $\sim 5.65\cdot 10^{21}$ s$^{-1}$}, respectively, (see Table~\ref{Tab:Hybrid.Rates}). 
{This lies between the one measured by the Mars Express-ASPERA instrument of $\sim 1.6\cdot 10^{23}$ s$^{-1}$ \citep{Barabash2007} and the one derived by MAVEN measurements of $\sim 5\cdot 10^{24}$ s$^{-1}$ \citep{jakosky18}. However, it is lower than the ones retrieved from other simulations for the present day such as the O$^+$ escape rates for a quiet and an active Sun in the study by \citet{fang13}, and the retrieved O$^+$ loss rates of \citet{chassefiere04}, \citet{chassefiere11}, and \citet{dong18} which are at the lower end of the \citet{fang13} values or even lower} (see also Fig.~\ref{Fig:EscapeRatesOxygen}). For C$^+$ {(see Fig.~\ref{Fig:EscapeRatesCarbon}), our escape rates {are lower by almost two orders of magnitudes than} the values of \citet{chassefiere04}, and \citet{chassefiere11}. {One reason for this discrepancy might be the low mixing ratio of C in the upper atmosphere within our simulations, since little of the CO$_2$ gets dissociated for 1\,EUV$_{\odot}$.}}


\vspace{1mm}

{For the 3\,EUV$_{\odot}$ case, the vast majority of the cold ions have energies of several eV (with an average of {$\sim 6.4$} and {$\sim 25.3$} eV for O$^+$ and C$^+$, respectively), while the hot ions have energies up to a few keV (with an average of {$\sim 8.3$} and {$\sim 11.1$ keV} for hot O$^+$ and hot C$^+$, respectively); the energy distributions of the precipitating planetary O$^+$ and C$^+$ ions at the lower boundary for all 3 EUV cases can be found in the supplement}. The total O$^+$ ion escape rate of {$\sim 2.2\cdot 10^{26}$\,s$^{-1}$} is {about the same} than the corresponding photochemical escape rate of $\sim 2.1\cdot 10^{26}$\,s$^{-1}$, as obtained in \citet{amerstorfer17}. The total C$^+$ escape rate of {$\sim 2.1 \cdot 10^{25}$\,s$^{-1}$} is {almost five times lower than} those of the suprathermal C atoms \citep{amerstorfer17} at that time. According to our simulations and in agreement with the results presented in \citet{dong18}, ion escape becomes the dominant loss process at Mars when its atmosphere is exposed to fluxes {$>$3 EUV$_{\odot}$, even though we receive higher values for 3\,EUV$_{\odot}$ and slightly lower ones for 10\,EUV$_{\odot}$ than \citet{dong18}}. The main reason {for the generally higher ion escape in the past} is related to the expansion of the upper atmosphere, which results in a larger planetary cross section for the incoming solar wind plasma and a larger ionization rate. Suprathermal atoms produced, e.g., by dissociative recombination have to move through {a denser background atmosphere from their main production region to the extended exobase level, leading to the hot particles colliding more often and thus being more easily thermalized, which results} in a lower total escape rate.
At present, the distance between the main production region of suprathermal atoms and the exobase level are close to each other so that many hot atoms with energies higher than the escape energy can reach the exobase. Therefore, escape of suprathermal atoms is the dominant non-thermal atmospheric escape process for heavy particles on present Mars.

For 10\,EUV$_{\odot}$, our total escape rate of O$^{+}$ ions reaches {$\sim 4.1\times 10^{26}$\,s$^{-1}$}, which {is slightly below but} in good agreement with the value obtained by \citet{dong18} of $\sim 1.1\times 10^{27}$\,s$^{-1}$. In comparison to \citet{boesswetter10}, who also modeled the escape of O$^+$ and CO$_2^+$ over the last 4.5 Ga by means of a 3D hybrid model, our loss rates and those of \citet{dong18} for 10\,EUV$_{\odot}$ are significantly lower. The reason for this discrepancy is their use of upper atmosphere profiles corresponding to present atmospheric CO$_2$ mixing ratios, which results in much higher IR-cooling and less expansion, while our profiles and those of \citet{dong18} are based on more accurate thermosphere models. In addition, the solar wind velocities and densities used by \citet{boesswetter10} are significantly higher but outdated. The same atmospheric profiles as in \citet{boesswetter10} (i.e. assuming no dissociation of the CO$_2$ molecules) were also used by \citet{Terada2009} and \citet{Sakata2020} and are therefore not accurate \citep{tian09,Scherf2021}.


\begin{table}[h!]
\caption{O$^+$ and C$^+$ production, impact and escape rates, and the production ratios in the three analysed EUV cases.}\vspace{2mm}
\label{Tab:Hybrid.Rates}
\centering \boldmath\bfseries
 \begin{tabular}{|c|l|p{.15\textwidth}|p{.15\textwidth}|p{.15\textwidth}|p{.2\textwidth}|}\hline
  EUV level & Species & Production\par rate [1/s] & Impact rate [1/s] & Escape rate [1/s] & Escape/ Production ratio\\\hline
 1 & Cold O$^+$ & $5.09\cdot 10^{24}$ & $4.75\cdot 10^{24}$ & $3.45\cdot 10^{23}$ & 0.07 \\
    & Hot  O$^+$ & $6.43\cdot 10^{23}$ & $2.84\cdot 10^{22}$ & $6.15\cdot 10^{23}$ & 0.96 \\\hline
    &Total O$^+$ & & & $9.60\cdot 10^{23}$ & \\\hline
   & Cold C$^+$ & $2.76\cdot 10^{22}$ & $2.57\cdot 10^{22}$ & $1.92\cdot 10^{21}$ & 0.07 \\
   & Hot  C$^+$ & $3.87\cdot 10^{21}$ & $1.46\cdot 10^{20}$ & $3.73\cdot 10^{21}$ & 0.96 \\\hline
   & Total C$^+$ & & & $5.65\cdot 10^{21}$ & \\\hline\hline
 3 & Cold O$^+$ & $2.63\cdot 10^{26}$ & $5.19\cdot 10^{25}$ & $2.11\cdot 10^{26}$ & 0.80 \\
    & Hot  O$^+$ & $5.34\cdot 10^{24}$ & $1.22\cdot 10^{23}$ & $5.22\cdot 10^{24}$ & 0.98 \\\hline
    &Total O$^+$ & & & $2.16\cdot 10^{26}$ & \\\hline
   & Cold C$^+$ & $1.06\cdot 10^{25}$ & $2.29\cdot 10^{24}$ & $8.35\cdot 10^{24}$ & 0.79 \\
   & Hot  C$^+$ & $1.27\cdot 10^{25}$ & $1.26\cdot 10^{23}$ & $1.26\cdot 10^{25}$ & 0.99 \\\hline
   & Total C$^+$ & & & $2.10\cdot 10^{25}$ & \\\hline\hline
10 & Cold O$^+$ & $3.62\cdot 10^{26}$ & $1.18\cdot 10^{25}$ & $3.50\cdot 10^{26}$ & 0.97 \\
   & Hot  O$^+$ & $6.43\cdot 10^{25}$ & $8.51\cdot 10^{23}$ & $6.35\cdot 10^{25}$ & 0.99 \\\hline
   &Total O$^+$ & & & $4.14\cdot 10^{26}$ & \\\hline
   & Cold C$^+$ & $4.36\cdot 10^{26}$ & $1.33\cdot 10^{25}$ & $4.23\cdot 10^{26}$ & 0.97 \\
   & Hot  C$^+$ & $9.53\cdot 10^{25}$ & $7.29\cdot 10^{23}$ & $9.46\cdot 10^{25}$ & 0.99 \\\hline
   & Total C$^+$ & & & $5.18\cdot 10^{26}$ & \\\hline
 \end{tabular}
\end{table}

\begin{table}[h!]
\caption{Sputter escape rates in the three analysed EUV cases for O, CO$_2$, CO, C, $^{36}$Ar and $^{38}$Ar.}\vspace{2mm}
\label{tab:Hybrid1}\boldmath\bfseries\footnotesize
\centering
 \begin{tabular}{|c|l|l|l|l|l|l|l|}\hline
EUV level & Incidend Ion & O & CO$_2$ & CO & C & Ar$^{36}$ & Ar$^{38}$\\\hline
 1 & Cold O  & $5.42\cdot 10^{24}$   & $1.59\cdot 10^{24}$    & $6.08\cdot 10^{23}$    & $7.74\cdot 10^{21}$    & $1.31\cdot 10^{20}$   & $2.09\cdot 10^{19}$\\

   & Cold C   & $2.60\cdot 10^{22}$   & $6.95\cdot 10^{21}$    & $2.81\cdot 10^{21}$    & $1.34\cdot 10^{19}$    & $5.92\cdot 10^{17}$   & $9.34\cdot 10^{16}$\\

   & Hot O   & $4.44\cdot 10^{22}$    & $1.32\cdot 10^{22}$    & $5.00\cdot 10^{21}$    & $2.22\cdot 10^{19}$    & $1.08\cdot 10^{18}$   & $1.71\cdot 10^{17}$\\

   & Hot C   & $1.78\cdot 10^{20}$    & $4.97\cdot 10^{19}$    & $1.94\cdot 10^{19}$    & $9.00\cdot 10^{16}$    & $4.14\cdot 10^{15}$  & $6.50\cdot 10^{14}$\\\hline

   & Total    & $5.49\cdot 10^{24}$    & $1.61\cdot 10^{24}$    & $6.16\cdot 10^{23}$    & $2.78\cdot 10^{21}$    & $1.33\cdot 10^{20}$  & $2.12\cdot 10^{19}$\\\hline\hline

 3 & Cold O & $1.77\cdot 10^{25}$ & $5.89\cdot 10^{22}$ & $4.75\cdot 10^{23}$ & $1.78\cdot 10^{23}$ & $6.19\cdot 10^{19}$ & $7.81\cdot 10^{18}$ \\

   & Cold C & $5.80\cdot 10^{23}$ & $1.81\cdot 10^{21}$ & $1.48\cdot 10^{22}$ & $6.15\cdot 10^{21}$ & $1.92\cdot 10^{18}$ & $2.41\cdot 10^{17}$ \\

   & Hot O  & $1.87\cdot 10^{23}$ & $1.82\cdot 10^{21}$ & $8.97\cdot 10^{21}$ & $1.52\cdot 10^{21}$ & $1.53\cdot 10^{18}$ & $1.99\cdot 10^{17}$ \\

   & Hot C  & $1.57\cdot 10^{23}$ & $1.65\cdot 10^{21}$ & $7.90\cdot 10^{21}$ & $1.25\cdot 10^{21}$ & $1.37\cdot 10^{18}$ & $1.77\cdot 10^{17}$ \\\hline

   & Total  & $1.86\cdot 10^{25}$ & $6.42\cdot 10^{22}$ & $5.07\cdot 10^{23}$ & $1.87\cdot 10^{23}$ & $6.67\cdot 10^{19}$ & $8.43\cdot 10^{18}$\\\hline\hline

10& Cold O  & $5.40\cdot 10^{24}$ & $7.03\cdot 10^{20}$ & $7.35\cdot 10^{22}$ & $1.07\cdot 10^{24}$ & $4.10\cdot 10^{19}$ & $4.44\cdot 10^{18}$ \\

  & Cold C  & $5.22\cdot 10^{24}$ & $6.81\cdot 10^{20}$ & $7.13\cdot 10^{22}$ & $1.09\cdot 10^{24}$ & $3.90\cdot 10^{19}$ & $4.33\cdot 10^{18}$ \\

  & Hot O   & $1.17\cdot 10^{24}$ & $2.49\cdot 10^{20}$ & $2.12\cdot 10^{22}$ & $2.03\cdot 10^{23}$ & $1.32\cdot 10^{19}$ & $1.47\cdot 10^{18}$ \\

  & Hot C   & $8.50\cdot 10^{23}$ & $1.74\cdot 10^{20}$ & $1.50\cdot 10^{22}$ & $1.49\cdot 10^{23}$ & $9.27\cdot 10^{18}$ & $1.02\cdot 10^{18}$ \\\hline

  & Total   & $1.26\cdot 10^{25}$ & $1.81\cdot 10^{21}$ & $1.81\cdot 10^{23}$ & $2.51\cdot 10^{24}$ & $1.02\cdot 10^{20}$ & $1.13\cdot 10^{19}$\\\hline
 \end{tabular}
\end{table}

%

By using the impact rates of exospheric O$^+$ and C$^+$ ions we also calculate the corresponding sputter escape rates for O, CO$_2$, CO, C, $^{36}$Ar, and $^{38}$Ar. Table~\ref{tab:Hybrid1} shows the escape rates of these elements for the 1, 3, and 10 EUV$_{\odot}$ cases; the calculated yields and the various parameters used for obtaining the sputter rates can be found in the supplement material.

{Figures~\ref{Fig:EscapeRatesOxygen} and \ref{Fig:EscapeRatesCarbon}} summarize our simulation outcomes for {O (Fig.~\ref{Fig:EscapeRatesOxygen}) and C (Fig.~\ref{Fig:EscapeRatesCarbon})}, {ion escape (panels b) and sputtering (panels c)} together with the results for suprathermal escape from \citet{amerstorfer17} (panels a) for all three EUV cases. The figure includes loss rates of several other earlier studies, i.e. hot oxygen loss through dissociative recombination (DR), ion escape and sputtering by \citet{chassefiere04} and \citet{chassefiere11}, ion escape and sputtering by \citet{feng13}, ion escape and suprathermal escape by \citet{dong18}, {measured O ion escape by Mars Express-Aspera \citep{Barabash2007} and derived O ion escape from Maven measurements \citep{jakosky18},} sputtering for present-day {based on MAVEN measurements} \citep{leblanc18}, and extrapolations for the loss of hot oxygen through DR by \citet{cravens17} and \citet{lillis17}. It is obvious, however, that extrapolations to higher EUV fluxes as applied by \citet{lillis17} through a simple power law fit can hardly yield appropriate escape rates, since the production of suprathermal atoms does not exponentially increase with increasing EUV flux, as outlined above and discussed in detail in \citet{amerstorfer17}.

For sputtering, {our theoretical result for C at present Mars is {by almost two orders} of magnitude lower as recently inferred by} \citet{leblanc18} and as those {compiled} by \citet{chassefiere04} and \citet{chassefiere11}. However, if we add-up the sputter escape rates of all C species, i.e., CO$_2$, CO, and C, then we obtain total C escape rates {that are about three times higher than in the study by \citet{leblanc18} for all C species together, which is due to our precipitation rate being higher. Furthermore, our CO$_2$ densities are much higher at 300\,km (i.e., $\sim 1.5 \times 10^{5}\rm cm^{-3}$ vs. $\sim 5 \times 10^{3}\rm cm^{-3}$ in the study by \citet{leblanc18}), which is well in agreement with the simulated density values obtained by \citet{Fox1997} of $\sim 5 \times 10^{5}\rm cm^{-3}$, and the measured values obtained by \citet{Mahaffy2015} of $\sim 3.5 \times 10^{5}\rm cm^{-3}$.}

Our oxygen sputter escape rates for 1\,EUV$_{\odot}$ {($5.5\times10^{24}$\,s$^{-1}$)}, on the other hand, are {by an order of magnitude} higher than those ($4\times10^{23}$\,s$^{-1}$) estimated by \citet{leblanc18} and also {slightly} higher than those ($2\times10^{24}$\,s$^{-1}$) given by \citet{Wang2014,Wang2015}. As noted by \citet{leblanc18}, they would have found an escape rate of $1\times10^{24}$\,s$^{-1}$ if they had used a precipitation rate of $\approx 1 \times 6.5\times10^{24}$\,O$^+/$s$^{-1}$ and the analytical relation between precipitating flux and induced atmospheric escape of \citet{Wang2015}. {The actual lower value of \citet{leblanc18} is attributed to the additional energy absorption of N and N$_2$, which was not considered by \citet{Wang2015}. Since we also neglect N and N$_2$ in our calculations, it is more resonable to compare our oxygen sputter rates with those of \citet{Wang2014,Wang2015}, showing that our rates are {only about 2 to 3 times} higher. The main reason for this difference seems to be due to the fact that the analytical model results in higher yields for oxygen \citep{johnson94,johnson00} than the model of \citet{Wang2015}.}


Our retrieved total escape rates of O for higher EUV fluxes fit quite well with the results of \citet{chassefiere11}, and \citet{dong18}. In all these cases, ion escape becomes the dominant process for increased EUV fluxes. For C, our ion escape rates are {almost 2 to about 3 orders of magnitude higher} than those of \citet{chassefiere04} and \citet{chassefiere11}, respectively. The reason for this is simple, though. In our case, at 10 EUV almost all of the CO$_2$ is dissociated leading to strong escape of carbon atoms due to their lighter weight and the accompanying expansion of the upper atmosphere, which is itself related to the dissociation of the infrared-cooler CO$_2$. Due to this effect, Mars could have lost a significant amount of CO$_2$ over its history, particularly for very high EUV fluxes $\geq$10\,EUV$_{\odot}$.

{Finally, it has to be noted that the integrated loss to space over time might be even higher than retrieved from our simulation, since we have chosen quiet conditions for the solar flux as well as for the solar wind, thereby not taking into account extreme solar events such as flares and interplanetary coronal mass ejections. Even though these extreme events only play a minor role for atmospheric escape at Mars at present-day, they likely have been significantly more frequent in the distant past \citep[e.g.,][]{Airapetian2016,Odert2017,Kay2019}.}

\subsection{Atmospheric Evolution}
\begin{figure}[b]
\centering
\includegraphics[width=12cm,angle=0]{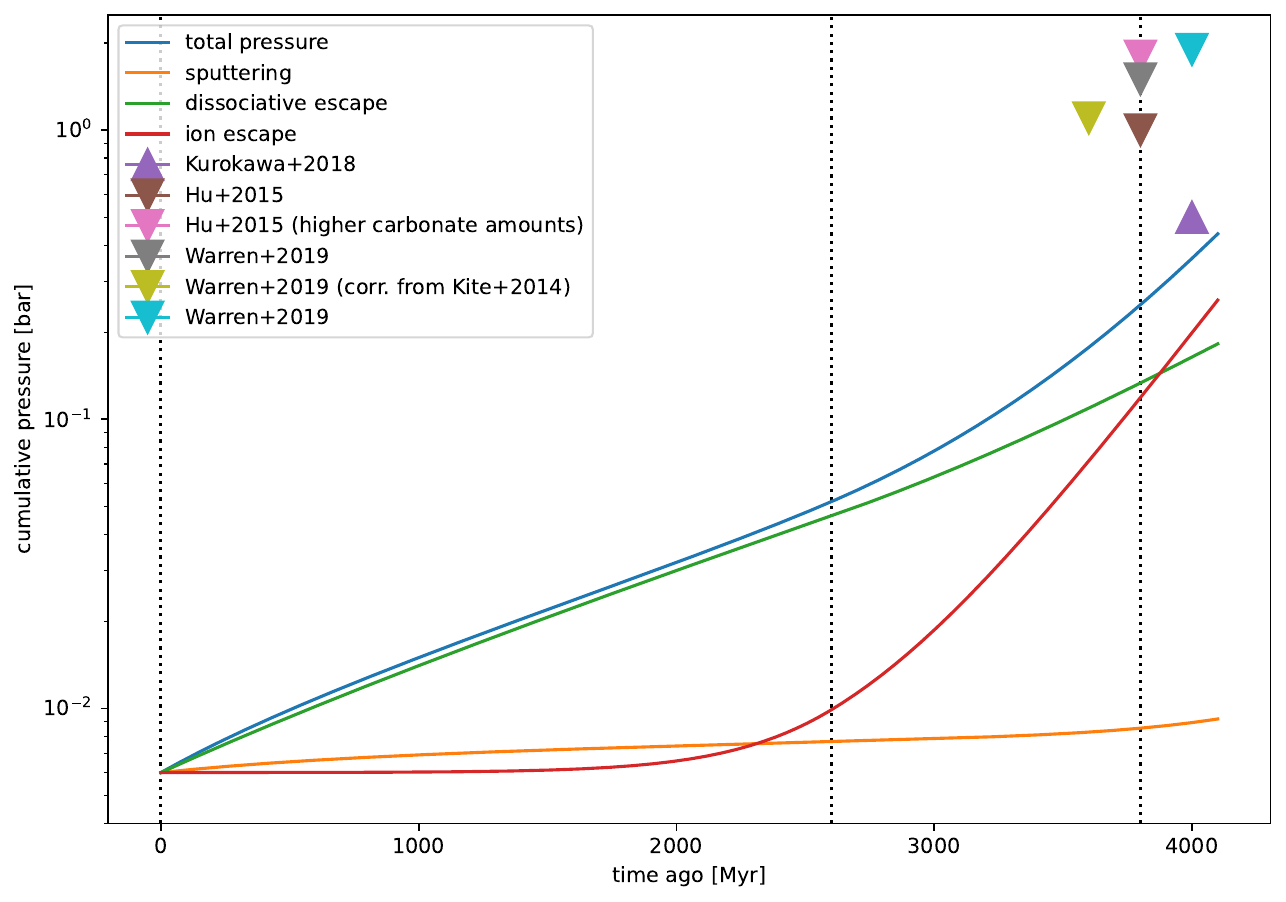}
\caption{Modeled CO$_2$-related surface pressure
evolution from present to 4.1 Ga caused by various atmospheric escape processes. The dissociative escape rates have been taken from \cite{amerstorfer17}, while ion escape and sputtering correspond to the results of this study. {The figure also shows maximum (down-pointing triangles) and minimum (up-pointing triangle) pressure estimates from the literature.}}
\label{Fig:CO2Pressure}
\end{figure}

Figure~\ref{Fig:CO2Pressure} {shows the integrated} CO$_2$ surface partial pressure evolution from today until 4.1 Ga, caused by the three main non-thermal atmospheric escape processes, namely suprathermal escape \citep{amerstorfer17}, as well as sputtering and ion escape as modeled in the present work. It is seen that sputtering induced escape does not contribute significantly to the total loss of CO$_2$, which is clearly dominated {by ion and suprathermal atom} escape for flux values $\geq$3 EUV$_{\odot}$. According to our results, more than {0.4\,bar} of CO$_2$ could have been  removed through the combination of the different non-thermal atmospheric escape processes over the last 4.1 Ga.
 Earlier than 4.1 Ga, the situation is getting more complex due to thermal escape becoming significant \citep[e.g.][]{tian09} and the presence of an intrinsic Martian magnetic field, which vanished around this time in the past \citep[e.g.][]{lillis13}. While it is yet not entirely clear, how such an intrinsic magnetic field might affect the efficiency of non-thermal escape processes \citep[being it a funnel or a shield, e.g.][]{gunell18,blackman18,egan19,Scherf2021}, it does not affect thermal escape, which could have been as high as  $\sim10^{30}$\,s$^{-1}$ for carbon at $\sim4.5$ Ga \citep{tian09}. {It also has to be noted that this surface pressure evolution does only include losses to space through thermal and non-thermal escape; carbonate formation as additional sinks may
allow for even higher pressures, since the total reservoir of CO$_2$ in carbonates at Mars is estimated to be up to $\approx 1\,$bar \citep[e.g.][]{Jakosky2019}. Similarly, atmospheric erosion through asteroid impacts could have provided another sink \citep[e.g.][]{pham16}. All of these sinks together, however, only provide a theoretical maximum pressure at a certain time that could have been lost until the present-day. The real pressure could have been significantly lower at any point in time \citep[see, e.g.,][for a discussion on this issue]{Scherf2021}.}

\vspace{1mm}

{To fit suprathermal, ion, and sputtering escape rates in between the three different EUV values, for each escape process a separate} exponential function in the form of
\begin{equation}\label{Eq:CO2evo}
[\mathrm{CO}_2]_{\rm esc}(t)=[\mathrm{CO}_2]_{\rm esc_0}\,e^{bt},
\end{equation}
was used \citep{chassefiere11}. For the periods between $1-3$ and $3-10$ EUV$_{\odot}$, the factor $b$ is determined by inserting $t=2.6$ and $t=3.8$ Ga, respectively, into Equation (\ref{Eq:CO2evo}), while for $t>3.8$ Ga, the escape was kept constant at the value of 10 EUV$_{\odot}$. This might be {a significant} underestimation, however, since the EUV flux should have been higher prior to 3.8\,Ga, also leading to higher non-thermal escape.

\begin{figure}[b]
\centering
\includegraphics[width=8cm,angle=0]{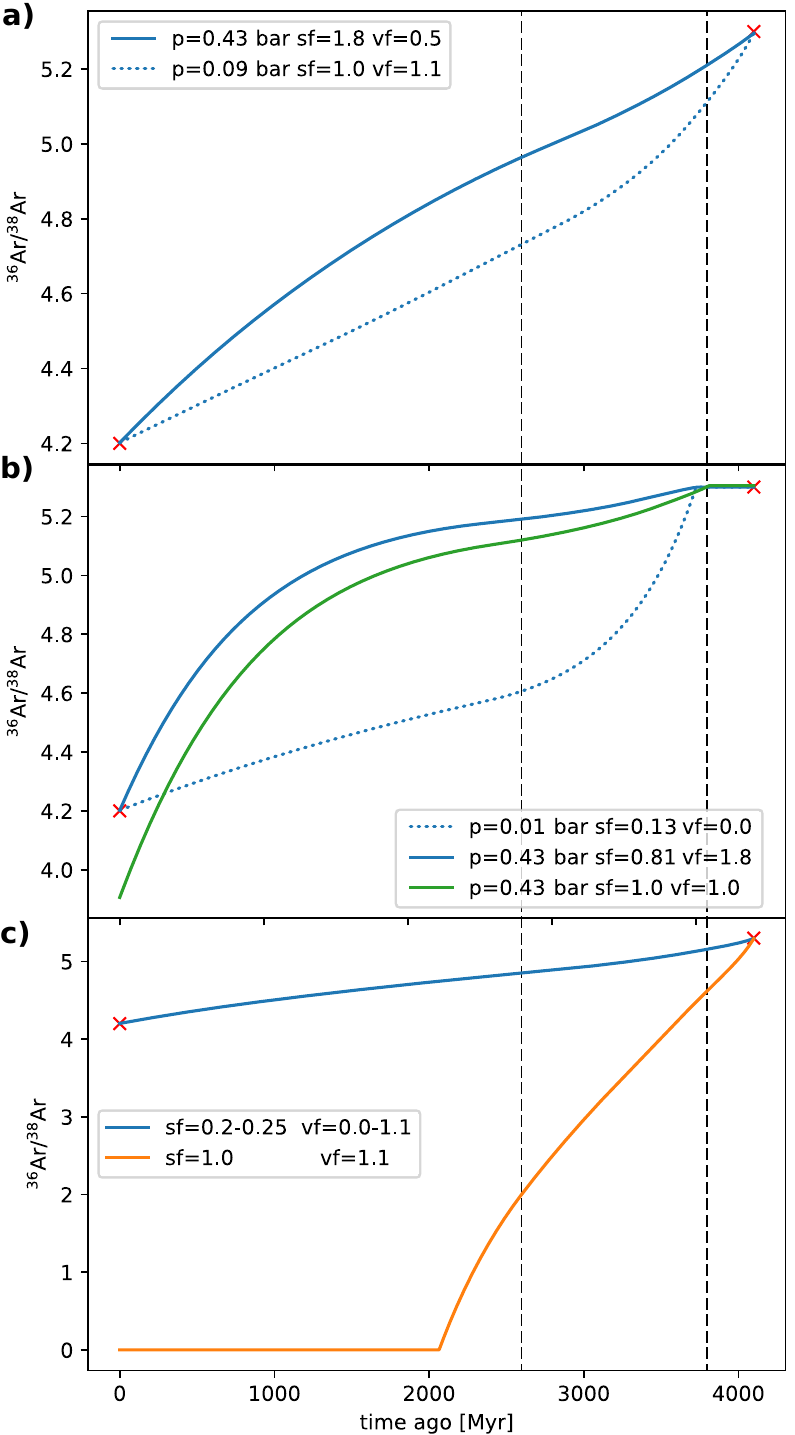}
\caption{Reproduction attempts from Mars' ALH 84001 based initial $^{36}$Ar/$^{38}$Ar ratio $\sim$ 4.1 Ga to the present atmospheric value with various initial CO$_2$-related surface pressure values and volcanic outgassing factors  $v_{\rm f}$.  Panel a) is derived with argon sputtering values for s$_{\mathrm{f}}=1$ taken from \citet{slipski16}, panel b) with the values derived from CO$_2$ sputtering as retrieved from our simulations. If one sets s$_{\mathrm{f}} = 1$ no solution can be found, whereas the $^{36}$Ar/$^{38}$Ar ratio can be reproduced for s$_{\mathrm{f}} = 0.16$ for pressures between 1 and 2.7\,bar. Panel c) is derived directly from our Ar sputtering values as given in Table~\ref{tab:Hybrid1}. The present-day ratio can only be reproduced if we reduce our sputter rates to  $s_\mathrm{f} = 0.16\,-\,0.18$, indicating that the total sputtering likely was lower than in our simulation. In this case no minimum and maximum CO$_2$ pressures can be derived.}
\label{Fig:Argon}
\end{figure}\vspace{1mm}

{Any change in the $^{36}$Ar/$^{38}$Ar ratio, on the other hand, is predominantly shaped by escape to space and replenishment through volcanic outgassing. }For its evolution and the therewith connected atmospheric escape over time, we compare the CO$_2$ sputter escape rates of \citet{slipski16} with our calculations. {Figure~\ref{Fig:Argon}a shows that, if one takes the argon escape rates of \citet{slipski16}, sputtering would have been able to create the present day $^{36}$Ar/$^{38}$Ar value from its initial ratio as preserved within ALH~84001 for an initial pressure range at 4.1 Ga of 0.1\,--\,0.45 bar for $s_{\rm f} = 1.0 - 1.8$ in Equation~(\ref{Eq:Lsput}), {if one takes about 0.45 bar as upper limit as retrieved from our escape study}. 

Figure~\ref{Fig:Argon}b shows the same, but with argon sputter values that are derived from our CO$_2$ sputtering {via Equation~\ref{Eq:Lsput}}. As can be seen, if keeping $s_{\rm f} = 1$ {(green line)}, an initial pressure of $\sim$0.43 bar cannot reproduce the present-day $^{36}$Ar/$^{38}$Ar ratio which might be an indication that our {CO$_2$ sputter rates} are too high. If one sets {$s_{\rm f} = 0.13\,-\,0.81$}, however, one can reproduce today's $^{36}$Ar/$^{38}$Ar ratio {for 0.01\,--\,0.45 bar which is similar to the pressure range that we retrieved with the values of \citet{slipski16}}.

Figure~\ref{Fig:Argon}c illustrates the evolution of the $^{36}$Ar/$^{38}$Ar ratio by taking our derived argon sputter rates calculated via Equation~\ref{eq:yield1} as shown in Table~\ref{tab:Hybrid1}. In this case, no minimum and maximum CO$_2$ pressure can be derived since Ar sputtering is not directly scaled with the evolution of CO$_2$. Moreover, the present-day ratio of $^{36}$Ar/$^{38}$Ar can, again, only be reproduced if we reduce our sputter ratios to a narrow range of {$s_{\rm f}\,=\,0.1\,-\,0.15$}. This indicates that the {Ar sputtering rates} that we derive from our simulation are} too high. {If we assume $s_{\mathrm{f}}=1.0$, all atmospheric $^{36}$Ar will escape {within $\approx$2\,Gyr}, as can be clearly seen in Figure~\ref{Fig:Argon}c (orange line). For values of {$s_{\rm f}<0.1$}, on the other hand, the fractionation would be too weak.

The {potential} overestimation of the sputter escape rates can have several reasons. 
One could be that we did not include $^{40}$Ar, the product from the radioactive decay of $^{40}$K, into our sputter calculations which might have reduced the sputter rates of $^{36}$Ar and $^{38}$Ar, since some of the precipitating ions could have also sputtered $^{40}$Ar instead of one of the other isotopes. For including $^{40}$Ar into the simulation one has to implement the whole radioactive K-Ar system into the model, which is beyond the scope of our present study; however, we are planning to include this in our future work since it might be a crucial factor for reconstructing the evolution of $^{36}$Ar/$^{38}$Ar.} Moreover, one should also note that especially for these extended early atmospheres another reason for some overestimation of the Argon isotope sputtering rates may result from the analytical model approach. {For finding and minimizing potential inherent errors in our calculation, future studies might, therefore, apply a combined upper atmosphere solar wind interaction Monte Carlo sputtering model to compare its outcome with the analytical approach that we have chosen in our study.}

{However, at least for the CO$_2$ sputter rates there might be another reason that makes our rates look too high.} The loss of $\sim$0.4 bar of CO$_2$ as displayed in Figure~\ref{Fig:CO2Pressure} is based on the assumption that CO$_2$ is removed from the atmosphere only via escape to space. However, since  CO$_2$ may also be stored in the surface, the real pressure might have been higher. 
From current atmospheric $^{13}$C/$^{12}$C, rock and soil carbonate measurements, \cite{hu15} constrained the atmospheric surface pressure of Mars at $\sim3.8$ Ga
to $<\,1$\,bar. Only scenarios with large amounts of carbonate deposition in open-water systems would permit higher values of up to 1.8 bar \citep{hu15}. {About 0.5\,bar for their nominal scenario with carbonate amounts based on present surface measurements, and up to 1.4\,bar in case of hidden carbonate reservoirs could have been sequestered into the soil. About ~0.5\,bar of CO$_2$ was lost to space from 3.8\,Ga until present according to their study which is slightly higher than our result. If one adds up the surface as a sink as investigated by \citet{hu15}, our result would allow up to 0.75 bar or in case of hidden carbonate reservoirs, even up to 1.65 bar of CO$_2$ at 3.8\,Ga, and 1.8\,bar at 4.1\,Ga.}

\begin{figure}[b]
\centering
\includegraphics[width=12cm,angle=0]{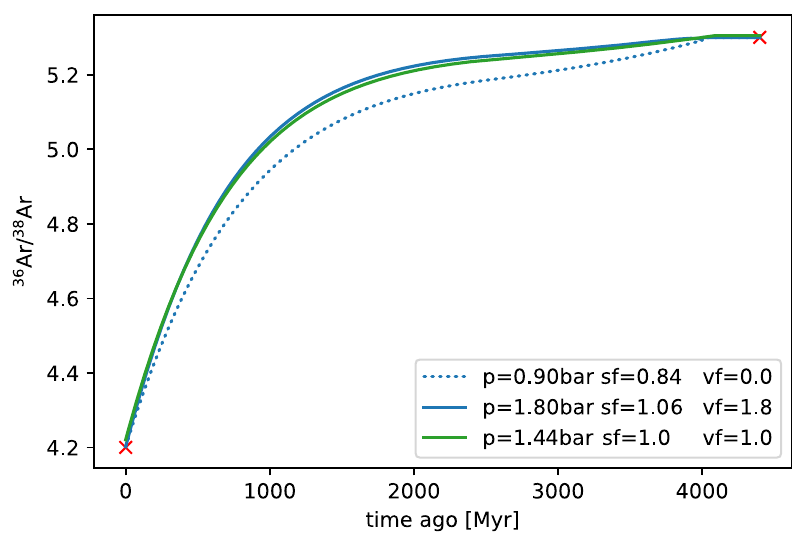}
\caption{{Same as Fig.~\ref{Fig:Argon}b, but with surface sinks, based on the results by \citet{hu15}, included into our reproduction attempt. By considering CO$_2$ deposited into the Martian soil, we can reproduce the present $^{36}$Ar/$^{38}$Ar ratio with an initial pressure at 4.1\,Ga between 0.90\,--\,1.80\,bar for $s_{\rm f} = 0.84 - 1.06$ with 1.44\,bar for $s_{\rm f} = 1$ and $v_{\rm f} = 1$.}}
\label{Fig:ArgonHu}
\end{figure}

{Moreover, since Equation~\ref{Eq:Lsput} depends on the initial CO$_2$ pressure, including deposition of the atmosphere into the ground might lead to higher values of $s_{\rm f}$ that are still possible to reproduce the present $^{36}$Ar/$^{38}$Ar ratio. By considering such a surface sink based on the results of \citet{hu15}, i.e., a decomposition of CO$_2$ of about 0.5\,bar since 3.8\,Ga for their nominal scenario, and up to 1.4\,bar in case of hidden carbonate reservoirs (see above), we can indeed reproduce the present argon fractionation even with $s_{\rm f} = 1$, as can be seen in Fig.~\ref{Fig:ArgonHu}. In such a case, the initial CO$_2$ surface pressure at 4.1\,Ga should have been between 0.9 and 1.8 bar for $s_{\rm f} = 0.84-1.06$, and 1.44\,bar for $s_{\rm f} = 1$, respectively. This is a strong indication that, instead of our CO$_2$ sputter rates being too high, atmosphere-surface interactions simply cannot be neglected.}

\vspace{1mm}

{Our result is also in close agreement with another recent study by \citet{Jakosky2019} which estimated the CO$_2$ inventory on Mars including its loss to space over time based on various previously published models. By reconstructing the Martian present-day $^{13}$C/$^{12}$C ratio of $\delta^{13}C \approx 45\permil$ \citep{mahaffy13,Webster2013} from an assumed initial value of -20 to -30$\permil$ based on carbon measurements within Martian meteorites \citep{Wright1986}, and by also taking into account the formation of carbonates which preferentially removes the heavier isotope from the atmosphere, \citet{Jakosky2019} found a lower limit of escape to space of about 1\,--\,2 bar CO$_2$ from 4.3\,Ga until present, {of which at least 0.7\,bar should have been lost prior to 3.8\,Ga}. This indicates that our result might also be in agreement with the evolution of the Martian $^{13}$C/$^{12}$C ratio over time, {at least for the lower limit of the total escape as estimated by \citet{Jakosky2019}.} However, simulating the evolution of $^{13}$C/$^{12}$C within our model would have been beyond the scope of this study, since other factors such as the formation of carbonates can also significantly alter the isotopic ratio of carbon.}

{Finally, there is another issue that has to be considered.} The measured value of $^{36}$Ar/$^{38}$Ar in the meteorite ALH84001 only shows that at $\sim 4.1$ Ga the argon isotope fractionation was 5.3. It does not indicate that sputtering immediately started to fractionate $^{36}$Ar from $^{38}$Ar. If the intrinsic magnetic field of Mars ceased later than at 4.1\,Ga, fractionation would have also started later, which could also be a possible scenario, since \citet{lillis13} found a time of cessation of 4.0\,-\,4.1\,Ga and \citet{milbury12} reported a later cessation time of even $\sim 3.6$ Ga.

\begin{figure}[b]
\centering
\includegraphics[width=12cm,angle=0]{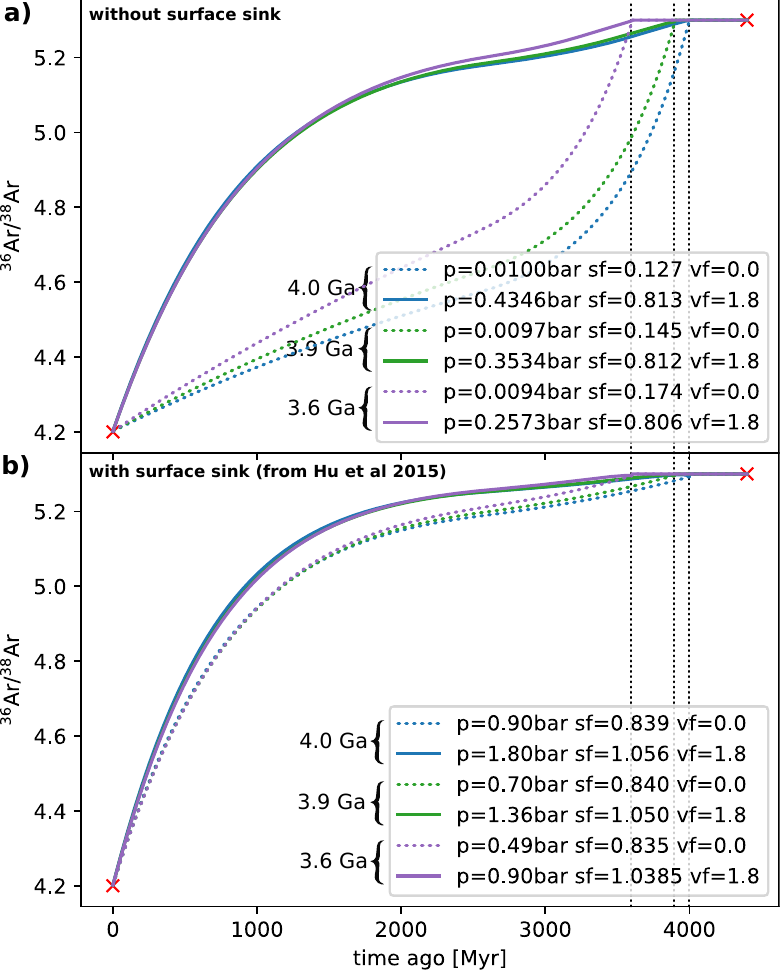}
\caption{a) The evolution of the $^{36}$Ar/$^{38}$Ar ratio as for case (b) of Fig.~\ref{Fig:Argon} but for different initial starting points of fractionation with our derived sputtering values. If fractionation started later, lower pressures are needed for reproducing $^{36}$Ar/$^{38}$Ar. {b) Same but with surface sinks included; see also Fig.~\ref{Fig:ArgonTimes}b.}}
\label{Fig:ArgonTimes}
\end{figure}

Figure~\ref{Fig:ArgonTimes}a shows the initial pressure range for different starting points of fractionation {based on atmospheric escape to space alone}. If fractionation started at 3.9\,Ga, the initial pressure {for $s_{\rm f} = 0.15\,-\,0.81$} at this particular time would have been between {0.01\,--\,0.35 bar}, and {for 3.6 Ga 0.01\,--\,0.26 bar {for $s_{\rm f} = 0.17\,-\,0.81$}}.

{Figure~\ref{Fig:ArgonTimes}b illustrates the initial pressure ranges for the same epochs, but with decomposition of CO$_2$ into the soil, as simulated by \citet{hu15}. In such a case significantly higher initial pressures would have been possible also for later ages, ranging from 0.9\,--\,1.8\, bar for 4.0, 0.7\,--\,1.4\,bar for 3.9, and 0.5\,--0.9\,bar for 3.6\,Ga, respectively. In all of these cases, $s_{\rm f}$ ranges between $\sim0.84-1.06$. The pressure values with surface sinks included also agree very well with all of the existing studies that derived upper or lower limits for the CO$_2$ pressure at early Mars (see Fig.~\ref{Fig:CO2Pressure}). Surface sinks clearly cannot be neglected when studying the evolution of the Martian atmosphere.}

\vspace{1mm}

Another possibility of a delayed on-set of Ar fractionation could be that Mars built up a CO$_2$ atmosphere only later than 4.1\,Ga with no dense atmosphere before. This might indeed be an additional option, since thermal escape is expected to have been very strong early-on during the pre-Noachian eon \citep{tian09}, leading to the potential escape of any early Martian CO$_2$-dominated atmosphere \citep{Scherf2021}, while argon might have been too heavy to escape thermally. Both, outgassing \citep[e.g.][]{grott11} and impact delivery \citep[e.g.][]{pham16} could have led to the built-up of at least a few hundred mbars of CO$_2$ after 4.1 Ga.

Keeping this in mind, it seems possible that the Martian atmosphere at 4.0 Ga was towards the lower range of potential pressure values that could reproduce $^{36}$Ar/$^{38}$Ar, {i.e. $0.9\,-\,1.8$ bar CO$_2$ {for $s_{\rm f} = 0.84\,-\,1.06$}}, or even below with fractionation starting later-on. In such a case, it is unlikely that early Mars had been a warm and wet planet, but might have been cold and try with some transient events of liquid water triggered by impacts or volcanic outgassing.

\section{Conclusion}\label{sec.Conclusion}

We simulated the ion escape of CO$_2$ related dissociation products C and O over the Martian history back until $\sim4.1$ Ga, the probably earliest time of the cessation of the Martian internal magnetic dynamo. The escape of cold and hot O$^+$ and C$^+$ was simulated for 1 EUV$_{\odot}$ (today), 3 EUV$_{\odot}$ ($\sim 2.6$ Ga), and 10 EUV$_{\odot}$ ($\sim3.8$ Ga) with a 3D global hybrid upper atmosphere model \citep{kallio03,dyadechkin13}, using the atmospheric profiles of \citet{tian09} and the hot O and hot C profiles of \citep{amerstorfer17}. We further investigated atmospheric sputtering of O, CO$_2$, CO, C, $^{36}$Ar and $^{38}$Ar caused by precipitating O$^+$ and C$^+$ ions calculated from our 3D hybrid simulations. We found that ion escape becomes the dominant loss process for solar fluxes $\ge$\,3 EUV$_{\odot}$ with loss rates at 10 EUV$_{\odot}$ of {$\sim 4.1\times10^{26}$\,s$^{-1}$} for O$^+$ and {$\sim 5.2\times10^{26}$\,s$^{-1}$} for C$^+$. For the same EUV flux, sputtering yields much lower escape rates of about {$1.3\times10^{25}$\,s$^{-1}$} for O and {$2.5\times10^{24}$\,s$^{-1}$} for C. If one extrapolates the total escape of CO$_2$ back in time until 4.1 Ga by assuming that any lost C atom originates from a CO$_2$ molecule, we find a maximum cumulative total CO$_2$ partial pressure {of more than 0.4\,bar that could have been lost through atmospheric escape. If one includes decomposition of CO$_2$ into the soil, this value could rise up to 1.8\,bar at 4.1\,Ga}

Moreover, we also studied the fractionation of argon isotopes from its initial chondritic value of $^{36}$Ar/$^{38}$Ar\,$\sim 5.3$ at 4.1 Ga through sputtering and volcanic outgassing until the present day. We found that, {if one includes the decomposition of CO$_2$ into the surface}, only CO$_2$ pressures within {0.9 - 1.8\,bar } at 4.1 Ga can reproduce the present day value, if argon fractionation by sputtering indeed started at 4.1 Ga, the time when the magnetic dynamo of Mars is believed to have faded away. Whether Mars could have had such a dense atmosphere at this time, however, is a matter of debate, since strong thermal escape prior to $\sim 4.0$ Ga \citep{tian09} might have prevented the built-up of such a dense early atmosphere.

However, the present-day ratio of $^{36}$Ar/$^{38}$Ar could also be reproduced if Mars outgassed a CO$_2$-dominated atmosphere with $\leq$\,1.4\,bar later than 4.1 Ga and/or if fractionation of argon through sputtering started $<4$ Ga. In this case, the required partial pressure needs to be {$\leq 0.7 - 1.8$\,bar}, while even later at 3.6 Ga it would have only been {$\leq 0.5 - 0.9$\,bar}. In case Mars only had a tenuous CO$_2$-dominated atmosphere during the pre-Noachian and only a few 100 mbar during the Noachian, it likely was a cold and dry body with only transient events of surficial liquid water. Therefore, further investigations based on isotopic fractionation, atmospheric escape, {but also atmosphere-surface interactions,} are needed in order to better constrain the climate conditions of early Mars and the subsequent evolution of its atmosphere. {This not only includes a simultaneous reconstruction of $^{13}$C/$^{12}$C over time together with $^{36}$Ar/$^{38}$Ar by taking into account the diverse processes that can affect both isotopic ratios, but, ultimately, also to explain the evolution of other volatile species at Mars besides CO$_2$ and Ar, such as H$_2$O and N within one comprehensive model.}

\section{Acknowlegements}
HIML, HL, and UVA acknowledge support by the Austrian Science Fundation via the project P 24247-N16. HL and SD acknowledge support by
the Austrian Science Fund (FWF) NFN project S11601-N16, “Pathways to Habitability: From Disks to Active Stars, Planets and Life” and the related FWF NFN subproject S11607-N16 ``Particle/Radiative Interactions with Upper Atmospheres of Planetary Bodies under Extreme Stellar Conditions''. The work of RJ was supported by the Academy of Finland (Decision No. 310444). Figures 2 and 3 were created using the VisIt open-source visualisation tool (Childs et al., 2012). We acknowledge Feng Tian from the Macau University of Technology and Science for giving support to this work. Finally, we thank B. Jakosky and an anonymous referee for their valuable comments which helped to improve the manuscript significantly.

\normalsize

\newpage

\bibliography{bibfile}

\end{document}


\section*{Supplement}

The first 12 figures illustrate the probability distribution of the precipitating O$^+$ and C$^+$ ions (cold and hot) that act as sputter agents for the respective lower boundaries of the 1, 3, and 10 EUV cases as a function of energy.

The subsequent 12 tables correspond to the respective sputter parameters and calculated yields for all cases and ions that are needed to calculate sputtering (see Equations in the main article in Section 4).

The final 3 tables give the neutral densities of the different species at the respective exobase levels for 1, 3, and 10 EUV.

\newpage

\begin{figure}
\centering
\includegraphics[width=12cm]{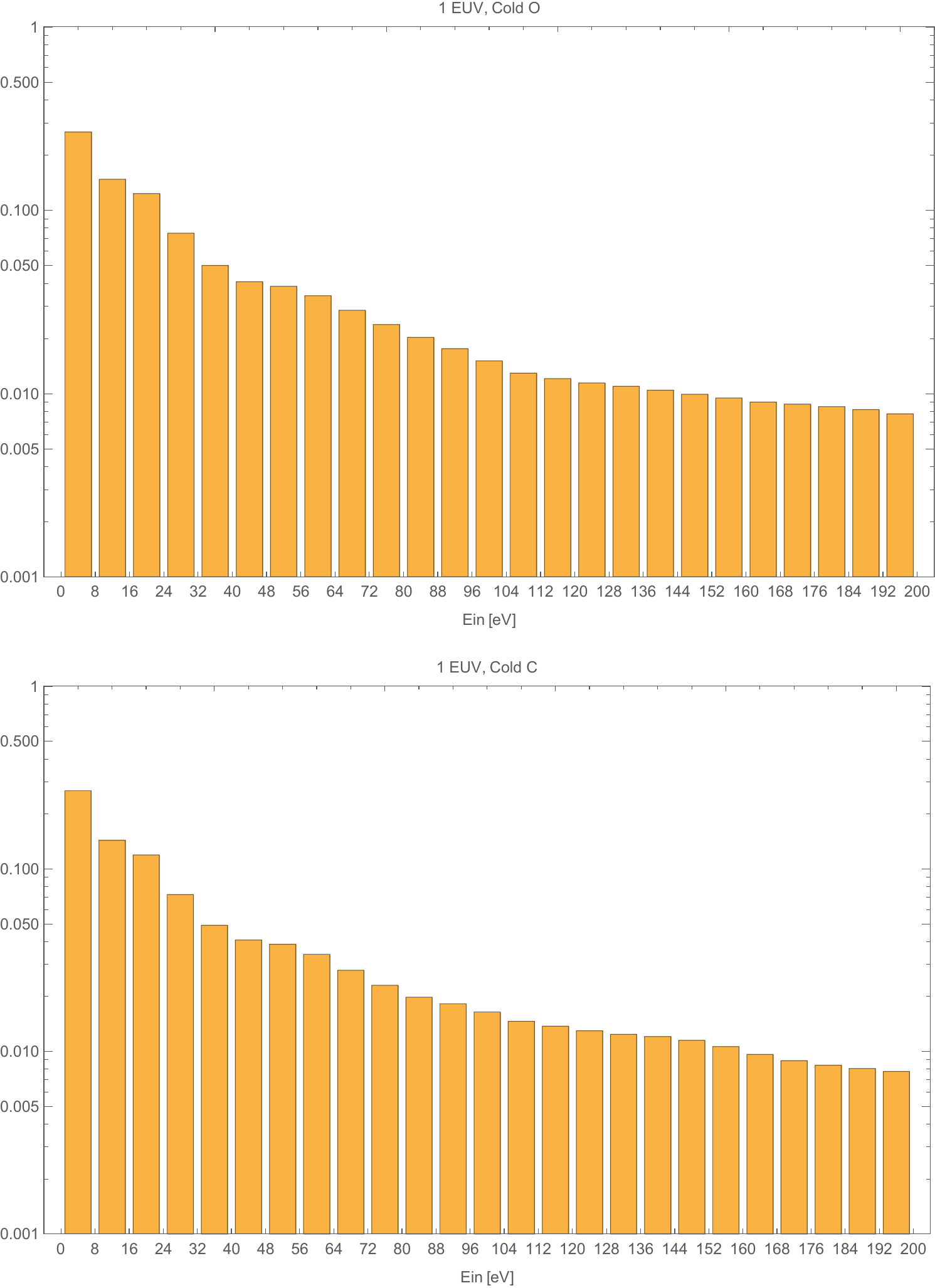}
\end{figure}

\begin{figure}
\centering
\includegraphics[width=12cm]{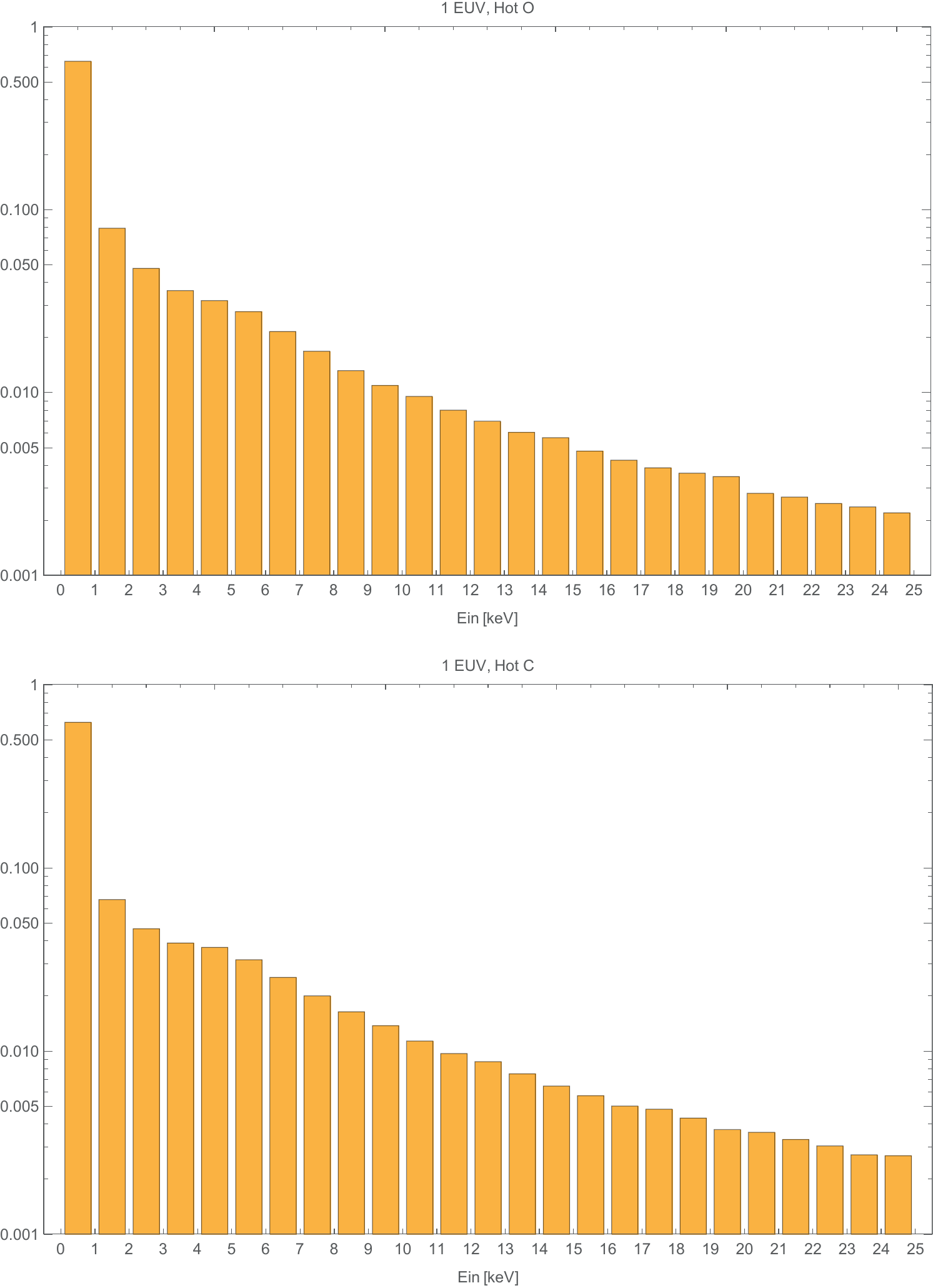}
\end{figure}

\begin{figure}
\centering
\includegraphics[width=12cm]{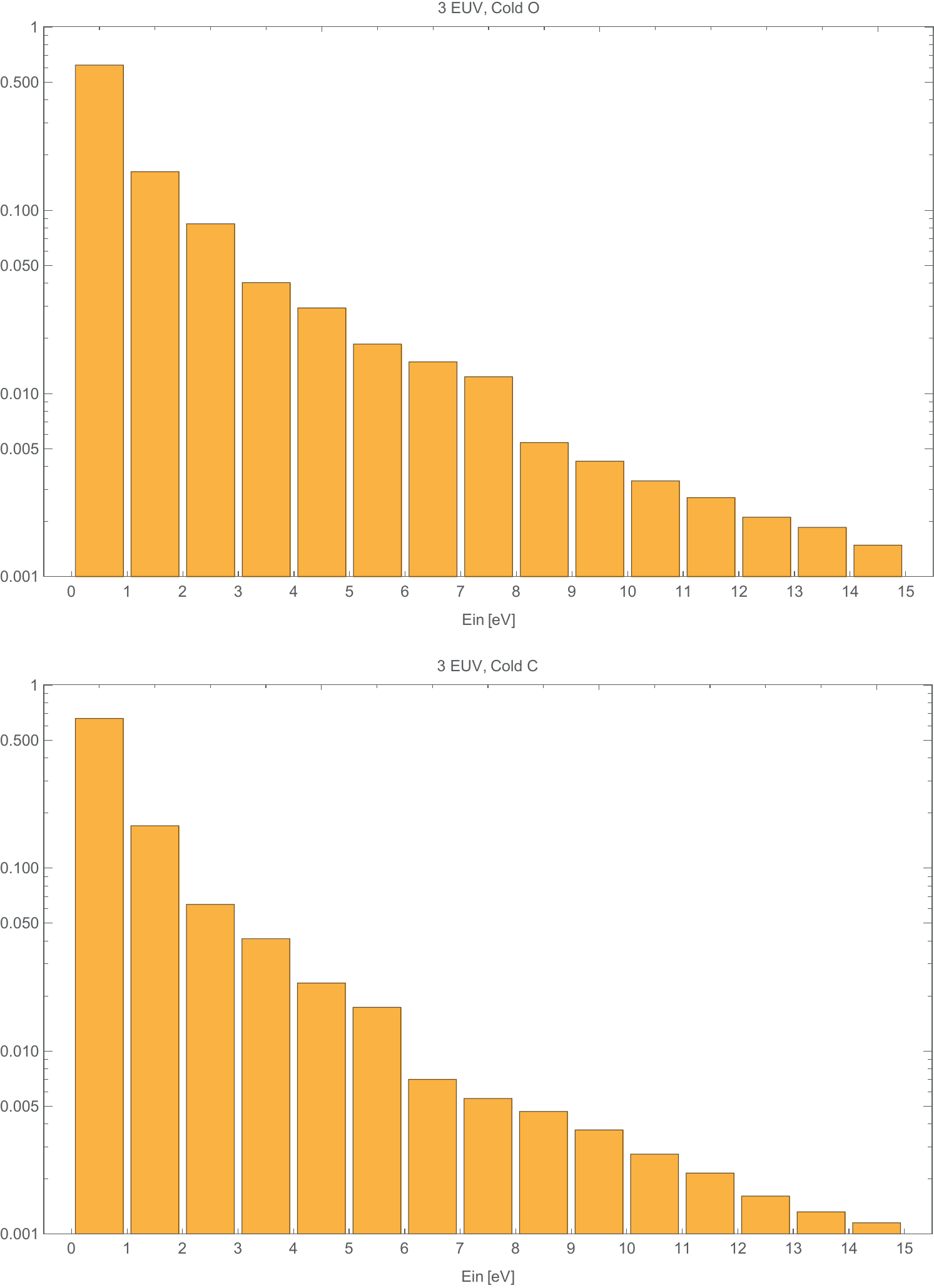}
\end{figure}

\begin{figure}
\centering
\includegraphics[width=12cm]{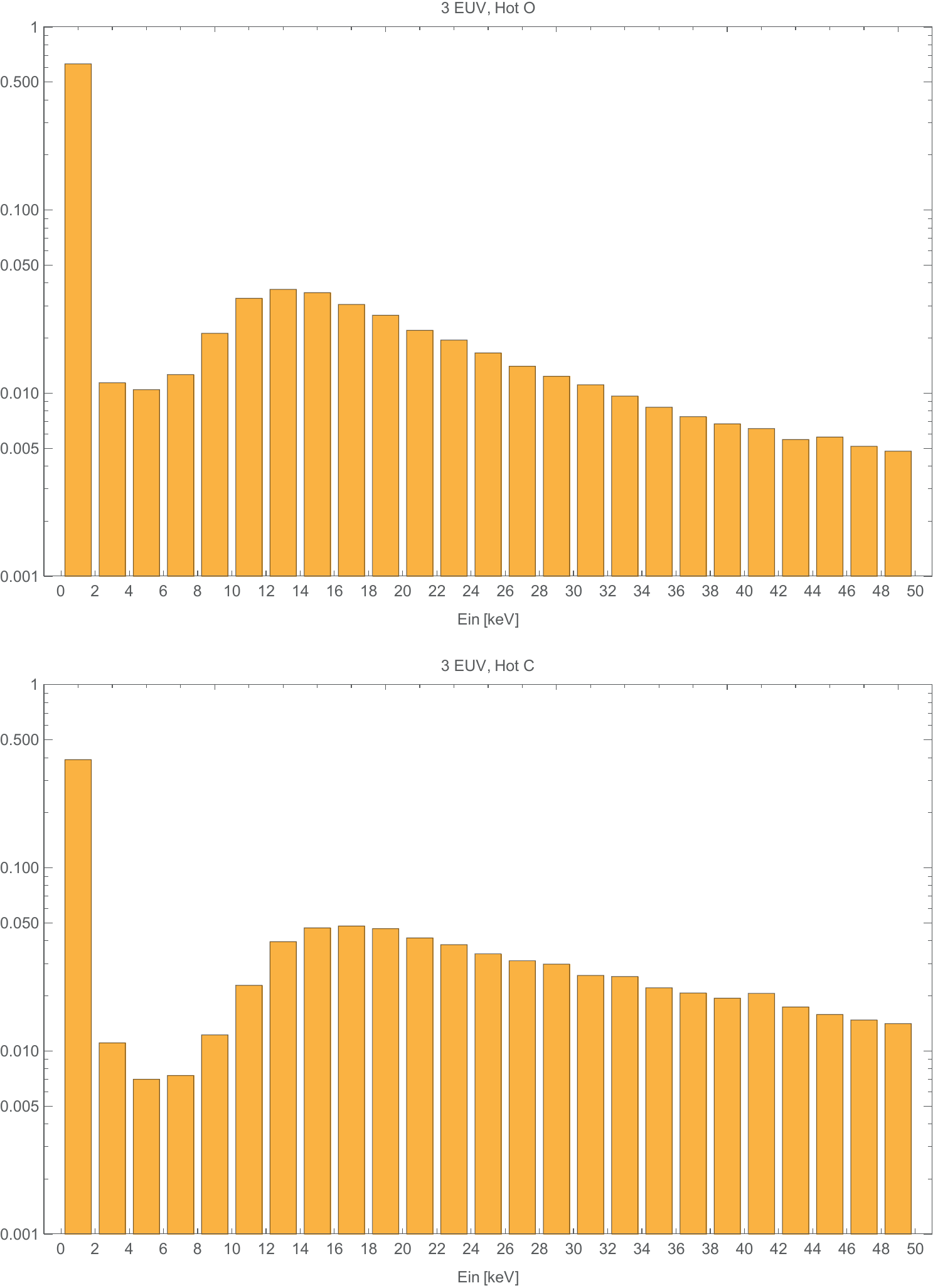}
\end{figure}

\begin{landscape}

\begin{figure}
\centering
\includegraphics[width=20cm,angle=0]{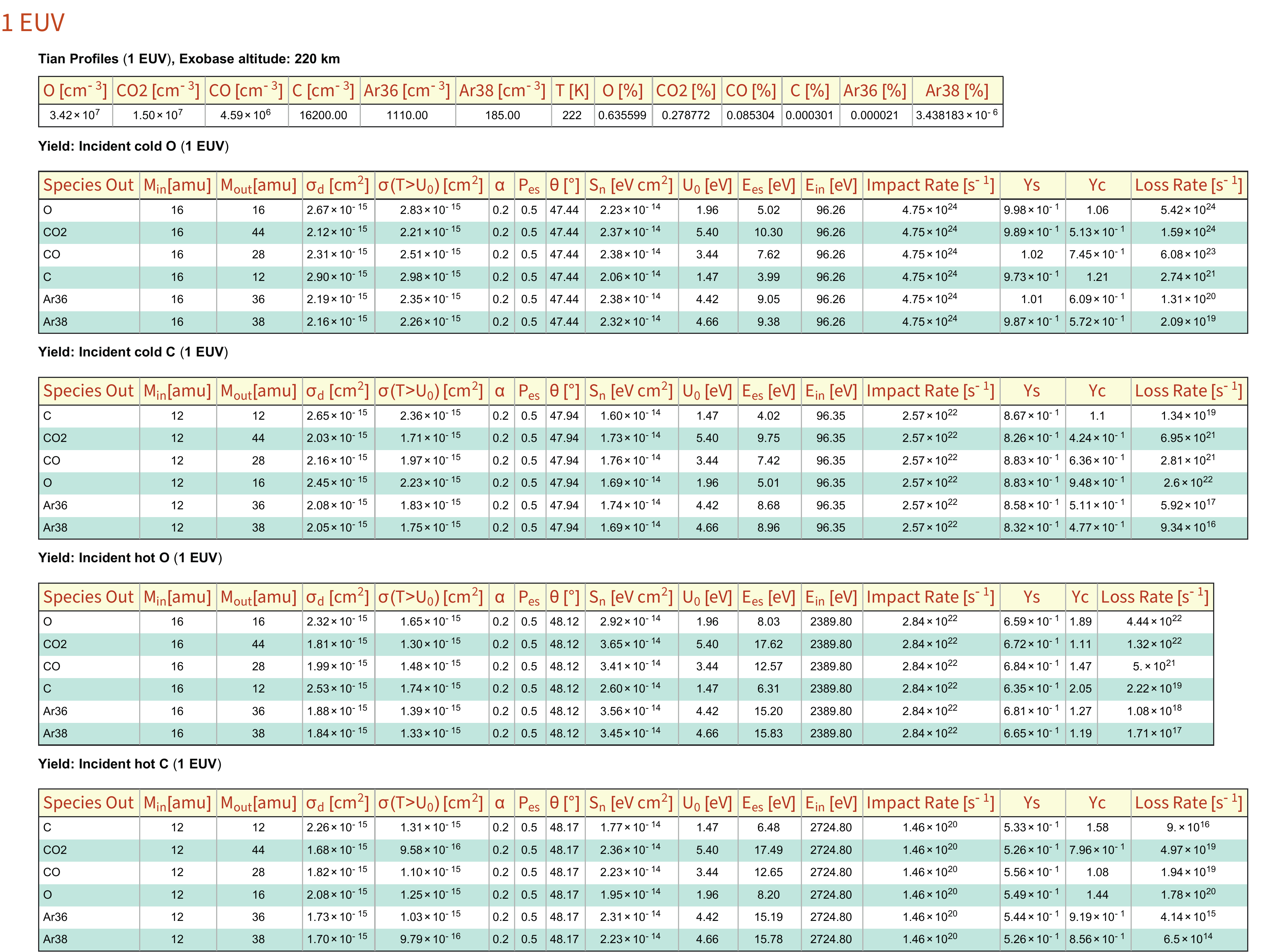}
\end{figure}

\begin{figure}
\centering
\includegraphics[width=20cm,angle=0]{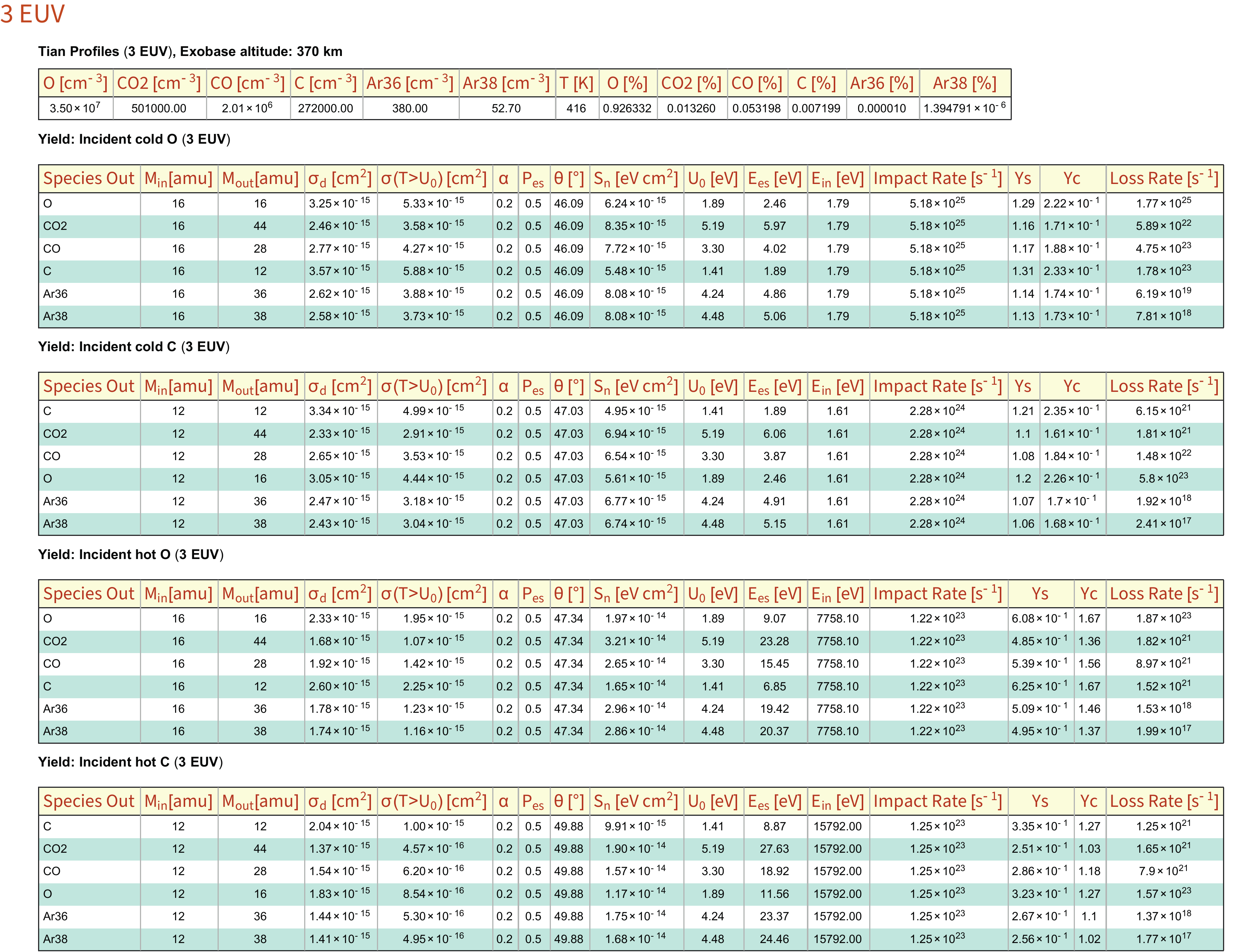}
\end{figure}

\begin{figure}
\centering
\includegraphics[width=20cm,angle=0]{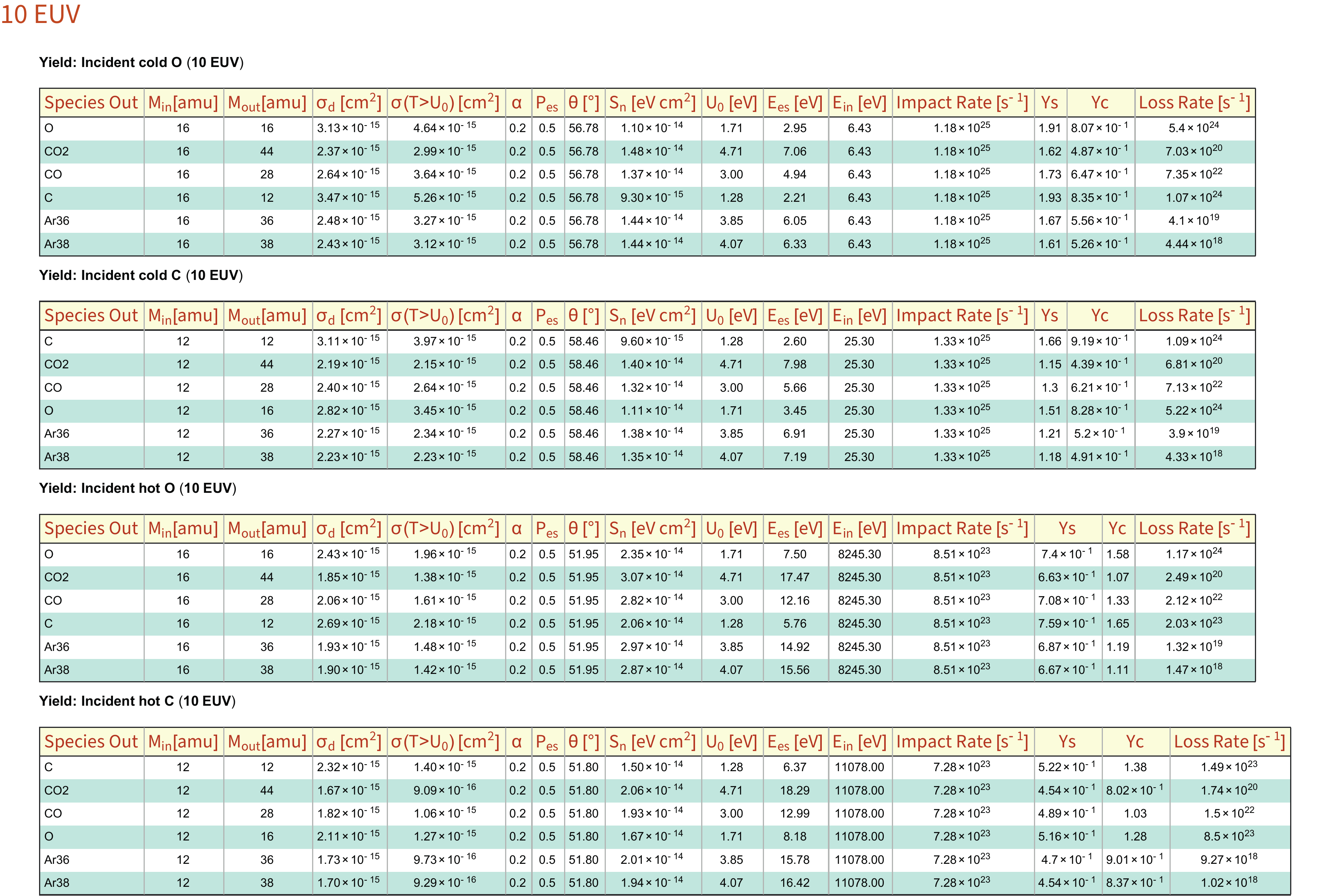}
\end{figure}

\end{landscape}